\documentclass[journal]{IEEEtran}
\usepackage{amsmath}
\usepackage{amsfonts}
\usepackage{cjk}
\usepackage{cite}
\usepackage{graphicx}
\usepackage{subfigure}
\usepackage{booktabs}
\usepackage{color}

\begin{document}

\title{A Simple Local Minimal Intensity Prior and An Improved Algorithm for Blind Image Deblurring}

\author{Fei Wen, 
        Rendong Ying,
        Yipeng Liu,
        Peilin Liu, 
        and Trieu-Kien Truong

\thanks{Copyright © 20xx IEEE. Personal use of this material is permitted. However, permission to use this material for any other purposes must be obtained from the IEEE by sending an email to pubs-permissions@ieee.org.}


\thanks{F. Wen, R. Ying, P. Liu and T.-K. Truong are with the Department of Electronic Engineering, Shanghai Jiao Tong University, Shanghai 200240, China (e-mail: wenfei@sjtu.edu.cn; rdying@sjtu.edu.cn; liupeilin@sjtu.edu.cn; truong@isu.edu.tw).}
\thanks{Y. Liu is with the School of Information and Communication Engineering, University of Electronic Science and Technology of China, Chengdu 611731, China (e-mail: yipengliu@uestc.edu.cn).}
}

\markboth{\underline{PUBLISHED IN  IEEE Transactions on Circuits and Systems for Video Technology, DOI:10.1109/TCSVT.2020.3034137}}
{Shell \MakeLowercase{\textit{et al.}}: Bare Demo of IEEEtran.cls for Journals}

\maketitle

\begin{abstract}
Blind image deblurring is a long standing challenging problem in image processing and low-level vision.
Recently, sophisticated priors such as dark channel prior, extreme channel prior, and local maximum gradient prior,
have shown promising effectiveness. However, these methods are computationally expensive.
Meanwhile, since these priors involved subproblems cannot be solved explicitly,
approximate solution is commonly used, which limits the best exploitation of their capability.
To address these problems, this work firstly proposes a simplified sparsity prior of local minimal pixels,
namely patch-wise minimal pixels (PMP).
The PMP of clear images is much more sparse than that of blurred ones,
and hence is very effective in discriminating between clear and blurred images.
Then, a novel algorithm is designed to efficiently exploit the sparsity of PMP in deblurring.
The new algorithm flexibly imposes sparsity inducing on the PMP under the maximum a posterior (MAP) framework rather than directly uses the half quadratic
splitting algorithm.
By this, it avoids non-rigorous approximation solution in existing algorithms, while being much more computationally efficient.
Extensive experiments demonstrate that the proposed algorithm can achieve better practical stability compared with state-of-the-arts.
In terms of deblurring quality, robustness and computational efficiency, the new algorithm is superior to state-of-the-arts.
Code for reproducing the results of the new method is available at https://github.com/FWen/deblur-pmp.git.
\end{abstract}


\IEEEpeerreviewmaketitle {}

\section{Introduction}
Blind image deblurring, also known as blind deconvolution,
aims to recover a sharp latent image from its blurred observation
when the blur kernel is unknown. It is a fundamental problem in
image processing and low level vision, which has been extensively
studied and is still a very active research topic in image processing and computer vision.

For single image deblurring and under the assumption that
the blur is uniform and spatially invariant,
the blur process can be modeled as a convolution operation \cite{19}, given by
\begin{equation}\label{1}
B = k \otimes I + n,
\end{equation}
where $B$, $k$, $I$, $n$, and $\otimes $ denote the blurred image, blur kernel, latent (clear) image,
additive noise, and convolution operator, respectively.
In the blind deblurring problem, only the blurred image $B$ is known {\textit{a prior}.
The objective} is to recover the kernel $k$ and the clear image $I$ simultaneously from $B$.
Basically, {it is a highly ill-posed problem as}
there exist many different solution pairs of $(k,I)$ giving rise to the same $B$.
Note that a typical undesired solution is that $I=B$ and $k$ being a delta blur kernel.

To make the blind deblurring problem well-posed, image prior and blur kernel model
exploitation is the key of most effective methods.
Well developed image priors include the gradient sparsity prior \cite{1,5,9,17,19},
normalized sparsity prior \cite{6}, patch prior \cite{11}, group sparsity prior \cite{46},
intensity prior \cite{14},
dark channel prior \cite{13,18}, extreme channel prior \cite{20}, latent structure prior \cite{19-struct-prior},
local maximum gradient prior \cite{47}, class-specific prior \cite{48},
and learned image prior using a deep network \cite{21}, to name just a few.
Meanwhile, blur kernel models include the non-uniform model with blur from multiple homographies \cite{22, 25, 45, 45-Zhang},
depth variation model \cite{26,27}, in-plane rotation model \cite{24}, and forward motion model \cite{23}.
Most of these methods exploit image prior and blur kernel model under the maximum a posterior (MAP) framework.
Generally, since the related deblurring problems are non-convex \cite{31},
incorporating regularization to exploit image prior and/or kernel model
helps to effectively increase the probability of achieving a good local solution.
In addition, heuristic edge selection is also an effect
way to help the MAP algorithms to avoid undesired trivial solutions.
For a more detailed discussion, see \cite{2,3}.

While image gradient sparsity is a popular and commonly used prior,
image intensity or gradient based priors have shown good complementary effectiveness
when jointly used with the image gradient prior \cite{13, 14, 20, 47}.
As priors and models designed for natural images usually
do not generalize well to specific images such as text images \cite{36},
face images, and low-light  images \cite{15},
simultaneously exploiting multiple priors
has the potential to achieve satisfactory performance on both natural and specific images \cite{13, 14, 20, 47}.

Though the sophisticated priors \cite{13, 20, 47} have shown promising effectiveness,
there exist two limitations:
\textit{i)} Jointly using multiple priors makes the corresponding algorithms computationally expensive.
\textit{ii)} Since these priors involved subproblems in the corresponding algorithms cannot be solved explicitly,
non-rigorous approximate solution is commonly used, which limits the best exploitation of the capability of such priors.
These limitations motivate us to develop a more effective and efficient method in this work.
The main contributions are as follows.

\subsection{Contribution}

Firstly, we propose a novel local intensity based prior,
namely the patch-wise minimal pixels (PMP) prior.
The PMP is a collection of local minimal pixels.
Intuitively, since the blur process has a smoothing effect on the image pixels,
the intensity of a local minimal pixel would increase after the blur process.
As a result, the PMP of clear images are much sparser than those of blurred ones.
The PMP metric is as simple as the direct intensity prior,
but is very effective in discriminating between clear and blurred images.
It can be viewed as a simplification of the dark channel prior in \cite{13},
which facilitates efficient computation while being effective.
A more detailed comparison with existing intensity priors \cite{13, 14} is provided in Section II-B.

Secondly, a novel algorithm is proposed to exploit the sparsity of PMP under the MAP framework.
The new algorithm flexibly imposes
sparsity {inducing} on the PMP of the latent image in the MAP deblurring process.
Compared with existing algorithms solving augmented MAP formulations directly based on half quadratic splitting, e.g., \cite{13, 20, 47},
the proposed algorithm greatly improves the computational efficiency in substance.
More importantly, while the algorithms \cite{13, 20, 47} use approximate solution in handling non-explicit subproblems,
the new algorithm avoids such approximation.
As a consequence, in comparison with state-of-the-art methods,
it is practically more robust and can achieve competitive performance on both natural and specific images.

Finally, extensive experimental results on blind image
deblurring  have been provided to evaluate the performance of the proposed method.
The results show that the proposed method can achieve state-of-the-art performance on both natural and specific images.
In terms of the deblurring quality, robustness and computational efficiency,
the proposed method is superior to the compared algorithms.

\subsection{Related Work}

In recent years, single image blind deblurring has made much progress
with the aid of various effective priors on images and blur kernels \cite{37}.
Most works are based on the variational Bayes and MAP frameworks \cite{1,2,3,4,5,6, 9, 10, 13, 14, 17,18,19,20,21, 35, 38,39,40}.
Typically, such a blind deblurring method generally has two steps.
First, blur kernel is estimated from the blurred observation under the MAP framework.
Then, based on the estimated blur kernel, the latent sharp image is
estimated via non-blind deconvolution methods, e.g., \cite{28,29,30}.

As the naive MAP method could fail on natural images, exploiting the statistical priors of natural images and selection of salient edges are the key of the success of state-of-the-art methods. The gradient sparsity prior of natural images is the most widely used prior. But it has been shown in \cite{5} that, in practice, the methods using the gradient sparsity prior in the MAP framework favor blurry images rather than clear ones. This limitation can be mitigated by techniques as explicit sharp edge pursuit \cite{1,3,35,41} or heuristic edge selection \cite{2}. However, the assumption of such techniques that there exist strong edges in the latent images may not always be satisfied.

There also exist various other image priors designed to reinvigorate MAP, such as normalized sparsity prior \cite{6}, internal patch recurrence \cite{10}, and direct intensity prior \cite{14}. Though effective for either natural or specific images, such priors usually cannot yield satisfactory performance on both natural and specific images. The recently proposed dark-channel prior \cite{13} and data driven learned prior \cite{21} can achieve satisfactory performance on both natural and specific images, but the involved optimization algorithms are computationally expensive.

Particularly, the intensity based priors considered in \cite{13, 14} are close to our proposed PMP prior.
As PMP is computed based on local minimal intensities,
it is fundamentally different from the intensity priors in \cite{13, 14}.
A detailed explanation on the difference is provided in Section II-B.
Furthermore, our algorithm exploits the PMP sparsity prior in a different way from that in \cite{13, 14} and,
as a consequence, it can reduce the computational complexity significantly while has more robust (stable) performance.
A detailed comparison on the algorithms is provided in Section IV.

Recently, data driven methods have also made much success with the aid of powerful deep learning techniques, e.g., \cite{24, 32,33,34, 43, 44}. For example, Sun \textit{et al.} \cite{24} endeavored to employ a convolutional neural network (CNN) to estimate and remove non-uniform motion blur. Nah et al. \cite{33} proposed a multi-scale CNN to recover the latent image in an end-to-end manner without any assumption on the blur kernel. Meanwhile, spatially variant recurrent neural network and scale-recurrent network have been designed for deblurring in \cite{32, 44}. Moreover, Kupyn \textit{et al.} \cite{43} proposed an end-to-end learned method for deblurring based on conditional generative adversarial networks (GAN). In addition, end-to-end CNN model for video deblurring has been considered in \cite{34}. Though these data driven methods can yield favorable performance in various scenarios, their success depends heavily on the consistency between the training data and the test data. This leads to the limitation of their generalization capability.

\textit{Outline:}
The rest of this paper is organized as follows.
Section II introduces the sparsity property of PMP and
discusses the connection between PMP and existing intensity priors.
The new algorithm is developed in detail in Section III.
Section IV presents comparison between the proposed algorithm and existing related algorithms.
Section V provides experimental results.
Finally, this paper concludes with a brief summery in Section VI.

\textit{Notations:}
$\otimes$ and $\nabla$ denote the convolution and gradient operation, respectively,
$\left\lceil  \cdot  \right\rceil $ denotes the ceil operator, $\circ $ stands for Hadamard (element-wise) product,
$\bar x$ denotes the conjugate of a complex quantity $x$.
$X(i,j)$ denotes the $(i,j)$-th element of a matrix $X$,
${\cal F}(X)$ denotes the 2-D FFT of $X$,
and ${\cal F}^{-1}(X)$ denotes the 2-D inverse FFT of $X$.

\section{Patch-Wise Minimal Pixels}

This section first introduces the proposed PMP prior
and presents analysis on its statistic property.
Then, comparison with existing intensity priors is provided.

\subsection{Patch-Wise Minimal Pixels}

{PMP is a collection of local minimal pixels over non-overlapping patches.}
Let an image $I \in {\mathbb{R}^{m \times n \times c}}$ be
divided into $P$ non-overlapped patches with a patch size of $r\times r$,
for which $P = \left\lceil {\frac{m}{r}} \right\rceil  \!\cdot\! \left\lceil {\frac{n}{r}} \right\rceil$.
The PMP is defined as
\begin{equation}\label{2}
{\cal{ P}}(I)(i) = \mathop {\min }\limits_{(x,y) \in {\Omega _i}} \left( {\mathop {\min }\limits_{c \in \{ r,g,b\} } I(x,y,c)} \right),
\end{equation}
for $i = 1,2, \cdots P$, where ${\Omega _i}$ denotes the index set
of the pixel locations of the $i$-th patch. It is easy to see that
${\cal P}(I) \in {\mathbb{R}^P}$ which contains patch-wise (local) minimal pixels of $I$.

\begin{figure}[!t]
\centering
{\includegraphics[width=7.5cm]{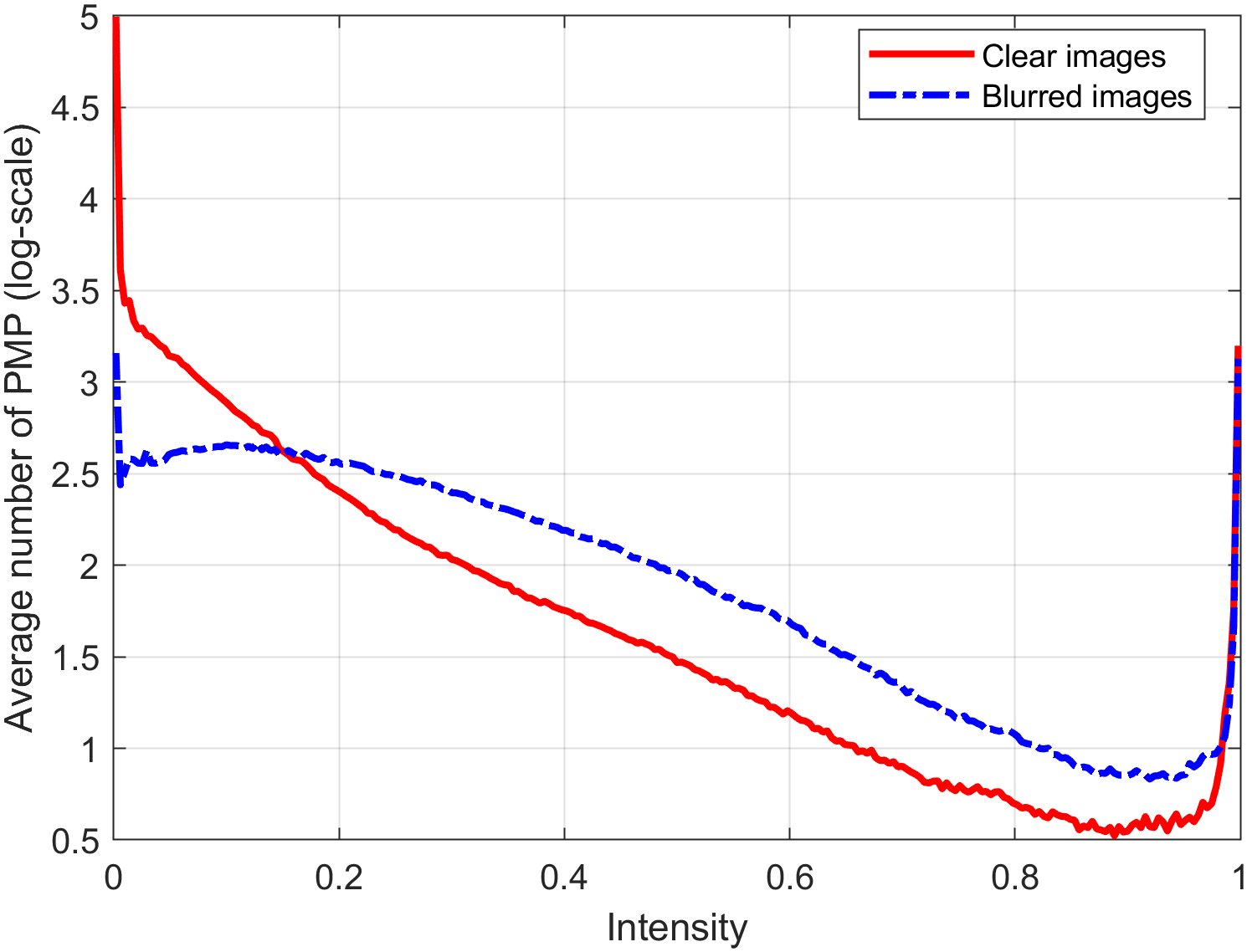}}
\caption{Intensity histogram for patch-wise minimal pixels of clear and blurred images over 5000 natural images (computed with an image patch size of $20\times20$). The PMP of clear images (under a threshold such as 0.9) follows a hyper Laplacian distribution and is much more sparse than the PMP of blurred images.}
\label{figure1}
\end{figure}

In what follows, we show that the PMP of blurred images are much less sparse than those of natural clear images.
Fig. \ref{figure1} compares the histogram statistic of PMP between clear images and their blurred counterparts
over more than 5000 natural images from the VGG\footnote{http://www.robots.ox.ac.uk/$\sim$vgg/data/} dataset.
The blurred images are synthesized from the clear ones using the blur kernels of the dataset \cite{5}.
It can be seen from Fig. \ref{figure1} that the PMP of clear natural images have significantly
more zero elements than those of blurred images. The PMP of clear images under a threshold (e.g., 0.9)
follow a hyper Laplacian distribution and manifest a sparsity property.
This sparsity property of PMP provides a natural metric to discriminate clear
images from blurred ones. Based on this property, the proposed algorithm imposes
sparsity {inducing} on PMP in the deblurring process to achieve more accurate kernel estimation.

Besides the empirical results shown in Fig. \ref{figure1},
the following result theoretically shows that blurred images
have less sparse PMP than their clear counterparts.

\textbf{Property 1:}
Let ${\cal P}(B)$ and ${\cal P}(I)$ denote the PMP of the blurred and clear images, respectively. Then
\begin{equation} \label{3}
{\cal P}(B) \ge {\cal P}(I).
\end{equation}

This property can be directly derived via extending Property 1 in \cite{13}.
It indicates that the blur process increases the values of PMP,
which gives rise to that the PMP of blurred images are less {sparse} than their clear counterparts.

In the following, without loss of generality, consider $c = 1$ for simplicity.
For the PMP ${\cal P}(I):{\mathbb{R}^{m \times n}} \to {\mathbb{R}^P}$ defined in (\ref{2}),
we further define its inverse operation for later use.
{Its inverse operation is defined by its transpose ${{\cal P}^T}(z):{\mathbb{R}^P} \to {\mathbb{R}^{m \times n}}$ for any $z \in {\mathbb{R}^P}$.}
Accordingly, we have
\begin{equation}\label{4}
{I_p}: = {{\cal P}^T}({\cal P}(I)) = I \circ M,
\end{equation}
where $M \in \mathbb{R}^{m \times n}$ is the binary mask corresponding to the PMP subset of $I$, with $M(i,j)=1$ if $I(i,j)$ is a minimal pixel of a patch and $M(i,j)=0$ otherwise.

\subsection{Comparison with Existing Intensity Sparsity Priors}

Intensity sparsity has also been considered in \cite{13, 14} for blind deblurring.
However, the proposed PMP is fundamentally different from
the intensity metrics considered in \cite{13, 14}, which is explained as follows.

The closest work exploiting intensity sparsity is the dark channel metric considered in \cite{13}.
For an image $I \in {\mathbb{R}^{m \times n \times c}}$, the dark channel is defined as
\begin{equation}\label{5}
{\cal D}(I)(i,j) = \mathop {\min }\limits_{(x,y) \in {\Omega _{i,j}}} \left( {\mathop {\min }\limits_{c \in \{ r,g,b\} } I(x,y,c)} \right),
\end{equation}
for $i = 1,2, \cdots m$ and $j = 1,2, \cdots n$, where ${\Omega _{i,j}}$ denotes
the index set of the pixel locations of the patch centered at the $(i,j)$-th pixel.
{It is easy to see that ${\cal D}(I) \in {\mathbb{R}^{m \times n}}$.}
While the dark-channel is computed in a convolution like manner with an output of size $m \times n$,
the proposed PMP is computed on non-overlapped patches with a vector output ${\cal P}(I) \in {\mathbb{R}^{P}}$ of size
$P = \left\lceil {\frac{m}{r}} \right\rceil  \!\cdot\! \left\lceil {\frac{n}{r}} \right\rceil$ for a patch size of $r \times r$.
As a result, the proposed PMP is much simpler than the dark-channel prior,
thereby facilitating the design of more efficient algorithm.

Moreover, the work \cite{14} uses sparsity {inducing}
directly on the intensity of the image for text image deblurring.
Since the intensity distribution of text images is close to two-tone,
using $\ell_0$-regularization to promote the intensity sparsity
has demonstrated outstanding effectiveness in text image deblurring.
However, the distribution of the intensity values of natural
images are more complex than that of text images,
and the direct intensity sparsity is not applicable to natural images.

In comparison, the proposed PMP metric is as simple as the direct intensity metric \cite{14},
but is very effective in discriminating between clear and blurred natural images as shown in Fig. \ref{figure1}.

\section{Proposed Deblurring Algorithm Using PMP Sparsity Regularization}

This section presents an efficient algorithm via flexibly incorporating the sparsity regularization of PMP into the conventional
MAP framework.
{The new algorithm is a variant of the half quadratic splitting algorithm,
but is different to the direct half quadratic splitting algorithm used in \cite{13, 14, 20, 47},
which is explained in detail in Section IV.}

Recall that the well-known MAP formulation is given by
\begin{equation}\label{6}
\mathop {\min }\limits_{k,I} {\cal L}(k \otimes I,B) + \gamma G(k) + \mu R(I),
\end{equation}
where $\gamma $ and $\mu$ are positive weight parameters,
and ${\cal L}$ is a data fidelity term,
which restricts $k \otimes I$ to be consistent with the blurred image $B$.
To make the problem well-posed,
$G$ and $R$ are penalty functions to exploit the structures
in the blur kernel and the latent image, respectively.

With the nature of that the gradient of natural images is sparse,
$R(I)$ is usually selected as the $\ell_0$-norm penalty of $\nabla I$ (the gradient of $I$).
Meanwhile, selecting both the loss function ${\cal L}$ and the penalty for the kernel as the ${\ell _2}$-norm yields
\begin{equation}\label{7}
\mathop {\min }\limits_{k,I} \left\| {k \otimes I - B} \right\|_2^2 + \gamma \left\| k \right\|_2^2 + \mu {\left\| {\nabla I} \right\|_0}.
\end{equation}
The ${\ell _2}$-norm is not only optimal for Gaussian noise,
but also enables the development of efficient algorithms
because it facilitates fast computation of
the involved subproblems via fast Fourier transform (FFT).

To further exploit the sparsity of the PMP of the latent image,
e.g., ${\cal P}(I) \in {\mathbb{R}^P}$ for a patch size of $r\times r$,
now consider a constrained extension of (\ref{7}) as
\begin{equation}\label{8}
\begin{split}
&\mathop {\min }\limits_{k,I} \left\| {k \otimes I - B} \right\|_2^2 + \gamma \left\| k \right\|_2^2 + \mu {\left\| {\nabla I} \right\|_0}\\
&\textrm{subject to  } ~{\cal P}(I)(i) \sim p(x),~~{\rm{for}}~ i \in \{ 1, \cdots ,P\}.
\end{split}
\end{equation}
As introduced in Section II, $p(x)$ is a probability density
function of a hyper Laplacian distribution for $x$ below a threshold such as 0.9.
The constrained problem (\ref{8}) is nonsmooth and nonconvex.
Similar to most existing deblurring algorithms in solving MAP-like objective functions,
we propose an efficient algorithm to solve (\ref{8})
via alternatingly update the blur kernel and the latent image.
In the proposed algorithm, the constraint in (\ref{8}) is approximately
imposed via iteratively sparsity {inducing} on ${\cal P}(I)$ in the latent image subproblem.

Note that a natural alternative of (\ref{8}) to promote sparsity of the PMP in the MAP framework is the formulation as follows:
\begin{equation}\label{9}
\mathop {\min }\limits_{k,I} \left\| {k \otimes I - B} \right\|_2^2 + \gamma \left\| k \right\|_2^2 + \mu {\left\| {\nabla I} \right\|_0} + \alpha {\left\| {{\cal P}(I)} \right\|_0},
\end{equation}
where $\alpha$ is a positive weight parameter and
the last term uses $\ell_0$-norm penalty to achieve sparsity {inducing} on the PMP of the latent image.
{This formulation can be solved directly by the half quadratic splitting algorithm
in an alternating manner similar to the algorithms in \cite{13, 14, 20, 47}.
However, we show that the proposed algorithm for solving (\ref{8})
is superior to the direct half quadratic splitting algorithm for solving (\ref{9}),
which will be explained in Section IV in detail.}

\subsection{Estimating the Latent Image}

Given an interim estimation of the blur kernel, denoted by $k^i$,
the latent image is updated via optimizing the following problem:
\begin{equation}\label{10}
\begin{split}
&\mathop {\min }\limits_I \left\| {{k^i} \otimes I - B} \right\|_2^2 + \mu {\left\| {\nabla I} \right\|_0}\\
&\textrm{subject to  } ~{\cal P}(I)(i) \sim p(x),~~{\rm{for}}~ i \in \{ 1, \cdots ,P\}.
\end{split}
\end{equation}
Using an auxiliary variable $G$ with respect to the image gradient $\nabla I$, the problem (\ref{10}) can be approximated by
\begin{equation}\label{11}
\begin{split}
&\mathop {\min }\limits_{I,G} \left\| {{k^i} \otimes I - B} \right\|_2^2 + \beta \left\| {\nabla I - G} \right\|_2^2 + \mu {\left\| G \right\|_0}\\
&\textrm{subject to  } ~{\cal P}(I)(i) \sim p(x),~~{\rm{for}}~ i \in \{ 1, \cdots ,P\},
\end{split}
\end{equation}
where $\beta > 0$ is a sufficient large penalty parameter such that
it enforces $\left\| {\nabla I - G} \right\|_2^2 \approx {\rm{0}}$, and hence $G \approx \nabla I$.

Without the constraint, such as in the case of the traditional MAP algorithm,
the problem (\ref{11}) can be typically solved using the block coordinate descent algorithm,
which alternatingly updates the two variables $I$ and $G$. The proposed algorithm also
solves (\ref{11}) via alternating between the variables $I$ and $G$
in which the constraint is approximately imposed.

Specifically, since the constraint in fact imposes a sparsity regularization on the PMP of $I$,
we use a simple thresholding/shrinkage step in the iteration procedure
to impose sparsity {inducing} on the PMP of $I$.
Given $I^t$ and at the $(t+1)$-th iteration of the latent image subproblem,
denote the PMP subset of $I^t$ by $I_{s}^t: = {\cal P}({I^t})$,
we iteratively impose thresholding on $I_s^t$ and
update $I$ and $G$ via the following steps.

First, let $\lambda > 0$ be a threshold parameter.
The PMP is thresholded as
\begin{equation}\label{12}
\begin{split}
&\tilde I_s^{t + 1,j}(i) = \left\{\! {\begin{array}{*{20}{l}}\!
{0,}&{|I_s^{t + 1,j}(i)| < \lambda }\!\\
\!{I_s^{t + 1,j}(i),}&{{\rm{otherwise}}}\!
\end{array}} \!\right.\!, \\
&~~{\rm{for}}~ i \in \{ 1, \cdots ,P\}.
\end{split}
\end{equation}
Subsequently, let ${\Omega ^{t + 1,j}}$ denote the index set of the PMP in ${I^{t + 1,j}}$
and define the mask corresponding to the PMP subset to be
\begin{equation}\label{13}
{M^{t + 1,j}}(i,j) = \left\{\! {\begin{array}{*{20}{l}}\!
{1,}&{{\rm{if  }}~(i,j) \in {\Omega ^{t + 1,j}}{\rm{ }}}\!\\
\!{0,}&{{\rm{otherwise}}}\!
\end{array}} \!\right.\!.
\end{equation}
Then, we update ${I^{t + 1,j}}$ as
\begin{equation}\label{14}
{\tilde I^{t + 1,j}} = {I^{t + 1,j}} \circ (1 - {M^{t + 1,j}}) + {{\cal P}^T}(\tilde I_s^{t + 1,j}),
\end{equation}
where ${{\cal P}^T}$ is the inverse operation of ${{\cal P}}$ defined in Section II-A.
With ${\tilde I^{t + 1,j}}$ given in (\ref{14}), the gradient-subproblem solves the following formulation
\begin{equation}\label{15}
{G^{t + 1,j + 1}} = \arg \mathop {\min }\limits_G \beta \left\| {\nabla {{\tilde I}^{t + 1,j}} - G} \right\|_2^2 + \mu {\left\| G \right\|_0},
\end{equation}
which is a proximal minimization and from \cite{42} the solution is explicitly given by
\begin{equation}\label{16}
\begin{split}
&{G^{t + 1,j + 1}}(i,j) = \left\{\! {\begin{array}{*{20}{l}}\!
{0,}&\!{{{(T(i,j))}^2} < {\mu}/{\beta} }\!\\
\!{T(i,j),}\!&\!{{\rm{otherwise}}}\!
\end{array}}\! \right.\!,\\
&~~{\rm{with}}~~ T = \nabla {\tilde I^{t + 1,j}}.
\end{split}
\end{equation}
Finally, the latent image is updated via solving the following problem
\begin{equation}\label{17}
{I^{t + 1,j + 1}} = \arg \mathop {\min }\limits_I \left\| {{k^i} \otimes I - B} \right\|_2^2 + \beta \left\| {\nabla I - {G^{t + 1,j + 1}}} \right\|_2^2,
\end{equation}
which can be efficiently computed by means of FFT as (\ref{18}) given in the next page,
where $\nabla = ({\nabla _h},{\nabla _v})$ and $G = ({G_h},{G_v})$ are used,
such that they correspond to image gradients in the horizontal and vertical directions, respectively.

\begin{figure*}[t]
\normalsize
\begin{equation}\label{18}
{I^{t + 1,j + 1}} =
{{\cal F}^{ - 1}}\left( {\frac{{\overline {{\cal F}({k^i})}  \circ {\cal F}(B) + \beta \left( {\overline {{\cal F}({\nabla _h})}  \circ {\cal F}(G_h^{t + 1,j + 1}) + \overline {{\cal F}({\nabla _v})}  \circ {\cal F}(G_v^{t + 1,j + 1})} \right)}}{{\overline {{\cal F}({k^i})}  \circ {\cal F}({k^i}) + \beta \left( {\overline {{\cal F}({\nabla _h})}  \circ {\cal F}({\nabla _h}) + \overline {{\cal F}({\nabla _v})}  \circ {\cal F}({\nabla _v})} \right)}}} \right).
\end{equation}
\end{figure*}

This algorithm is summarized in Algorithm 1, which contains two loops.
In Algorithm 1, $a$ is a positive increasing factor, which is set to $a=2$ in the experiments.
Extensive numerical studies show that the inner loop usually converges within a few iterations.
For example, we use $J=3$ in the experiments in Section V.

\begin{table}[!t]
\normalsize
\begin{tabular}{p{8.4cm}}
\toprule
\textbf{Algorithm 1:} Latent image estimation\\
\midrule
\hangafter 1
\hangindent 1.5em
\noindent
\textbf{Input:} Blurred image $B$, interim kernel estimation $k^i$.\\
$\beta  \leftarrow {\beta _0}$, ${I^0} \leftarrow B$.\\
\textbf{While} $\beta \le {\beta _{max}}$ \textbf{do} ($t=0,1,2,\cdots$) \\
~~~~${I^{t + 1,0}} \leftarrow {I^t}$.\\
~~~~\textbf{For} $j = 0:J - 1$ \textbf{do} \\
~~~~~~~~Compute the mask ${M^{t + 1,j}}$ via (\ref{13}) based on ${I^{t + 1,j}}$.\\
~~~~~~~Obtain $\tilde I_s^{t + 1,j}$ via (\ref{12}) and further update ${\tilde I^{t + 1,j}}$\\
~~~~~~~via (\ref{14}).\\
~~~~~~~Compute gradient thresholding to obtain ${G^{t + 1,j + 1}}$\\
~~~~~~~via (\ref{16}).\\
~~~~~~~Update ${I^{t + 1,j + 1}}$ via (\ref{18}).\\
~~~~\textbf{End for} \\
~~~~${I^{t + 1}} \leftarrow {I^{t + 1,J}}$.\\
~~~~${\beta} \leftarrow a{\beta }$.\\
\textbf{End while}\\
${I^{i + 1}} \leftarrow {I^{t + 1}}$.\\
\textbf{Output:} Intermediate latent image estimation ${I^{i + 1}}$.\\
\bottomrule
\end{tabular}
\end{table}

\subsection{Estimating the Blur Kernel}

Similar to other existing state-of-the-art algorithms, the kernel estimation
is performed in the gradient space. As it has been shown that
gradient space based methods are more accurate than intensity space based ones \cite{3, 9, 17}.
Specifically, given an interim estimation of the latent image,
denoted by $I^i$, the blur kernel is updated via solving
\begin{equation}\label{19}
{k^{i + 1}} = \arg \mathop {\min }\limits_k \left\| {k \otimes (\nabla {I^i}) - \nabla B} \right\|_2^2 + \gamma \left\| k \right\|_2^2.
\end{equation}
Due to its quadratic form, the solution can be efficiently computed by means of FFT, give by
\begin{equation}\label{20}
\begin{split}
&{k^{i + 1}} = \\
&{{\cal F}^{ - 1}}\left( {\frac{{\overline {{\cal F}({\nabla _h}{I^i})}  \circ {\cal F}({\nabla _h}B) + \overline {{\cal F}({\nabla _v}{I^i})}  \circ {\cal F}({\nabla _v}B)}}{{\overline {{\cal F}({\nabla _h}{I^i})}  \circ {\cal F}({\nabla _h}{I^i}) + \overline {{\cal F}({\nabla _v}{I^i})}  \circ {\cal F}({\nabla _v}{I^i}) + \gamma }}} \right).
\end{split}
\end{equation}
Moreover, the estimated kernel is further refined via setting the negative elements to zero and normalization.
In practical implementation, the multi-scale deconvolution scheme \cite{3}
is adopted to estimate the kernel in a coarse-to-fine manner.
The main steps for kernel estimation at a single scale level are shown in Algorithm 2.

\begin{table}[!t]
\normalsize
\begin{tabular}{p{8.4cm}}
\toprule
\textbf{Algorithm 2:} Blind blur kernel estimation\\
\midrule
\hangafter 1
\hangindent 1.5em
\noindent
\textbf{Input:} Blurred image $B$, kernel initialization $k^0$ from the estimation in the last coarser-scale.\\
\textbf{For} $i = 1:max\_iter$ \textbf{do}\\
~~~~Estimate ${I^i}$ via Algorithm 1 using ${k^{i - 1}}$.\\
~~~~Estimate ${k^i}$ via (\ref{20}).\\
\textbf{End for}\\
$\hat k \leftarrow {k^i}$, $\hat I \leftarrow {I^i}$.\\
\textbf{Output:} Kernel estimation $\hat k$, intermediate image $\hat I$.\\
\bottomrule
\end{tabular}
\end{table}

\subsection{Implementation Tricks}

To make the augmented objective function in (\ref{11}) accurately approaching that in (\ref{10}), a sufficiently large value of $\beta$ is desired, ideally $\beta  \to \infty $. However, with a very large value of $\beta$, an alternating algorithm directly minimizing (\ref{11}) would be very slow and impractical. To address this problem, a standard trick is to use a continuation process for $\beta$. In other words, one starts with a properly small value of $\beta$ and gradually increase it by iteration until a large target value is reached. This trick is used in Algorithm 1 with a $a > 1$.

The thresholding step of the PMP in (\ref{12}) corresponds to a nonconvex $\ell_0$-regularization. In addition, the objective function in (\ref{11}) is also nonconvex. Hence, with different initialization and/or parameter setting, a nonconvex algorithm would end up with one of its many local minimizers. In view of this, in implementing Algorithm 1, we use soft-thresholding instead of the hard-thresholding in the first few scales in the multi-scale procedure and then turn back to the hard-thresholding. As the soft-thresholding corresponds to the convex $\ell_1$-regularization, this ``first loose and then tight'' strategy makes the proposed algorithm more stable and performing satisfactorily.

\section{Comparison with the Half Quadratic Splitting Algorithm Solving the Regularized MAP Formulation (\ref{9})}

As mentioned in Section III, a natural alternative of (\ref{8}),
which can incorporate sparsity {inducing} of the PMP into the MAP framework, is given by (\ref{9}).
{Compared with the formulation (\ref{8}),
the formulation (\ref{9}) is even more explicit and can be solved by means of the half quadratic splitting algorithm.
However, in this section we show that, compared with the direct half quadratic splitting algorithm solving (\ref{9}),
the proposed algorithm is not only superior in computational complexity,
but also can avoid non-rigorous approximate solution in solving the regularized MAP problem.}

To solve (\ref{9}) with a given interim kernel estimation $k^i$, the latent image problem becomes
\begin{equation}\label{21}
\mathop {\min }\limits_I \left\| {{k^i} \otimes I - B} \right\|_2^2 + \mu {\left\| {\nabla I} \right\|_0} + \alpha {\left\| {{\cal P}(I)} \right\|_0}.
\end{equation}
Similar to [13], [14], using two auxiliary variables $G$ and $Z$ with respect to $\nabla I$ and ${\cal P}(I)$, respectively, the problem (21) is approximated by
\begin{equation}\label{22}
\begin{split}
\mathop {\min }\limits_I \left\| {{k^i} \otimes I - B} \right\|_2^2 + \mu {\left\| G \right\|_{\rm{0}}}+ \alpha {\left\| Z \right\|_{\rm{0}}} &+ \beta \left\| {\nabla I - G} \right\|_2^2\\
& + \rho \left\| {{\cal P}(I) - Z} \right\|_2^2,
\end{split}
\end{equation}
where $\beta$ and $\rho$ are positive penalty parameters.
Given ${I^t}$ at the $(t + 1)$-th iteration,
the $G$- and $Z$-subproblems are proximity operators,
which can be efficiently solved in an element-wise manner as
\begin{equation}\label{23}
\begin{split}
&{Z^{t + 1}}(i,j) = \!\left\{\! {\begin{array}{*{20}{l}}
\!{0,}&\!{{{(Y(i,j))}^2} < {\alpha}/{\rho} }\!\\
\!{Y(i,j),}&\!{{\rm{otherwise}}}\!
\end{array}} \!\right.\!,\\
&~~{\rm{with}}~~Y = {\cal P}({I^t}),
\end{split}
\end{equation}
and
\begin{equation}\label{24}
\begin{split}
&{G^{t + 1}}(i,j) =\! \left\{ \!{\begin{array}{*{20}{l}}
\!{0,}&\!{{{(T(i,j))}^2} < {\mu}/{\beta} }\!\\
\!{T(i,j),}&\!{{\rm{otherwise}}}\!
\end{array}} \!\right.\!,\\
&~~{\rm{with}}~~T = \nabla {I^t}.
\end{split}
\end{equation}
Then, the $I$-subproblem becomes
\begin{equation}\label{25}
\mathop {\min }\limits_I \left\| {{k^i} \otimes I - B} \right\|_2^2 + \beta \left\| {\nabla I - {G^{t + 1}}} \right\|_2^2 + \rho \left\| {{\cal P}(I) - {Z^{t + 1}}} \right\|_2^2.
\end{equation}
In view of that there exist two augmentation terms in the nonconvex problem (\ref{22}),
to make the algorithm practically working well,
a standard trick is to use a continuation process for each of $\beta $ and $\rho$
similar to the algorithms in \cite{13, 14}. In such a manner,
the main steps of the half quadratic splitting algorithm are sketched in Algorithm 3.

\begin{table}[!t]
\normalsize
\begin{tabular}{p{8.4cm}}
\toprule
\textbf{Algorithm 3:} Latent image estimation via solving (\ref{22})\\
\midrule
\hangafter 1
\hangindent 1.5em
\noindent
\textbf{Input:} Blurred image $B$, interim kernel estimation $k^i$.\\
$\rho  \leftarrow {\rho _0}$, ${I^0} \leftarrow B$.\\
\textbf{While} $\rho \le {\rho _{max}}$ \textbf{do} ($t=0,1,2,\cdots$) \\
~~~~Compute ${Z^{t + 1}}$ via (\ref{23}) using ${I^t}$.\\
~~~~$\beta  \leftarrow {\beta _0}$, ${I^{t + 1,0}} \leftarrow {I^t}$.\\
~~~~\textbf{While} $\beta \le {\beta _{max}}$ \textbf{do} ($j=0,1,2,\cdots$) \\
~~~~~~~~Obtain ${G^{t + 1,j + 1}}$ via (24) using ${I^{t + 1,j}}$.\\
~~~~~~~~Solve (25) to update ${I^{t + 1,j+1}}$.\\
~~~~~~~~${\beta } \leftarrow a{\beta }$.\\
~~~~\textbf{End while} \\
~~~~${I^{t + 1}} \leftarrow {I^{t + 1,J}}$.\\
~~~~${\rho } \leftarrow a{\rho}$.\\
\textbf{End while}\\
${I^{i + 1}} \leftarrow {I^{t + 1}}$.\\
\textbf{Output:} Intermediate latent image estimation ${I^{i + 1}}$.\\
\bottomrule
\end{tabular}
\end{table}

Although both Algorithms 1 and 3 contain two main loops,
the former is much more efficient than the latter in practice.
That is because the penalty parameters ${\rho _{max}}$ and ${\beta _{max}}$
in Algorithm 3 should be chosen sufficiently large to make (\ref{22}) accurately approximating for (\ref{21}),
while a small value of $J$ (e.g., $J=3$) is enough for Algorithm 1 to give satisfactory performance.

Moreover, although the $I$-step in Algorithm 3 solves a quadratic problem (\ref{25}),
it cannot be efficiently solved via FFT similar to (\ref{18}).
A strategy to explicitly solve the least-square problem (\ref{25}) in closed-form
is to vectorize the variables and convert the convolution operation into linear multiplication.
However, this least-square problem involves computing the inverse of high-dimensional matrices
of size $(mn) \times (mn)$ with $m \times n$ be the size of $I$.
Thus, it is computationally expensive to handle practical-sized inputs.
Meanwhile, since ${\cal P}(I)$ is a subsampling function of $I$ and
only contains a subset of the pixels of $I$, clearly, the problem (\ref{25})
cannot be solved in closed-form via FFT similar to (\ref{18}) and the algorithms as given in \cite{6, 7, 10}.

With the definition in (\ref{4}), problem (\ref{25}) can be equivalently rewritten as
\begin{equation}\label{26}
\begin{split}
\mathop {\min }\limits_I \left\| {{k^i} \otimes I - B} \right\|_2^2 &+ \beta \left\| {\nabla I - {G^{t + 1}}} \right\|_2^2 \\
&+ \rho \left\| {{I_p} - {{\cal P}^T}({Z^{t + 1}})} \right\|_2^2.
\end{split}
\end{equation}
Now, let ${{ \breve{I}} _p} = I \circ (1 - M)$ be the
complementary set of $I_p$ such that it satisfies ${I_p} + \breve{I}_p = I$.
It is easy to see from (\ref{25}) that $I_p$ and $\breve{I}_p$
are coupled through the kernel convolution operation.
With this in mind, to enable FFT based efficient solution,
we can iteratively solve (\ref{26}) via alternating between $I_p$ and $\breve{I}_p$.
For example, we first fix $\breve{I}_p$ to solve $I_p$ by
\begin{equation}\label{27}
\begin{split}
\mathop {\min }\limits_{{I_p}} \left\| {{k^i} \otimes {I_p} \!-\! (B - {k^i} \otimes {\breve{I} _p})} \right\|_2^2 + &\beta \left\| {\nabla {I_p} \!-\! ({G^{t + 1}} - \nabla {\breve{I} _p})} \right\|_2^2\\
 + &\rho \left\| {{I_p} - {{\cal P}^T}({Z^{t + 1}})} \right\|_2^2,
\end{split}
\end{equation}
and then fix $I_p$ to solve $\breve{I}_p$ by
\begin{equation}\label{28}
\mathop {\min }\limits_{{\breve{I}_p}} \left\| {{k^i} \otimes {\breve{I}_p} \!-\! (B - {k^i} \otimes {I_p})} \right\|_2^2 + \beta \left\| {\nabla {\breve{I}_p} \!-\! ({G^{t + 1}} \!-\! \nabla {I_p})} \right\|_2^2.
\end{equation}

Thanks to the quadratic form of (\ref{26}), iteratively solving (\ref{27}) and (\ref{28})
is guaranteed to converge to the global minimizer of (\ref{26}) with any bounded initialization.
Even though both (\ref{27}) and (\ref{28}) can be efficiently solved by means of FFT,
iteratively solving them makes Algorithm 3 having three iteration loops,
and hence results in an increase of computational complexity in terms of runtime.

Note that in the dark-channel based method \cite{13},
the $I$-subproblem has a similar formulation as (\ref{25}),
where ${\cal P}(I)$ is replaced by the dark-channel extraction operation,
which is solved via FFT in close-form. The close-form solution
is derived via implicitly using an approximation. That is
\begin{equation}\label{29}
\left\| {{{\cal D}_{{I^t}}}(I) - u} \right\|_2^2 \approx \left\| {I - \left( {{\cal D}_{{I^t}}^T(u) + \breve{I} _d^t} \right)} \right\|_2^2,
\end{equation}
where ${{\cal D}_{{I^t}}}$ denotes the dark-channel extraction operator
based on ${I^t}$, ${\cal D}_{{I^t}}^T$ is the inverse operator of ${{\cal D}_{{I^t}}}$,
and $\breve{I} _d^t$ is the complementary set of $I_d^t$
with $I_d^t$ being the subset pixels of $I^t$ which forms the dark-channel.
In fact, the right term in (\ref{29}) can be viewed as an approximation of
\begin{equation}\label{30}
\begin{split}
&\left\| {{\cal D}_{{I^t}}^T({{\cal D}_{{I^t}}}(I)) - {\cal D}_{{I^t}}^T(u)} \right\|_2^2\\
& = \left\| {{I_d} - {\cal D}_{{I^t}}^T(u)} \right\|_2^2=\left\| {I - \left( {{\cal D}_{{I^t}}^T(u) + \breve{I} _d} \right)} \right\|_2^2,
\end{split}
\end{equation}
where ${\cal D}_{{I^t}}^T({{\cal D}_{{I^t}}}(I)) = {I_d}$ and $I = {I_d} + \breve{I} _d$ are used.

Similar approximation has also been used in \cite{20,47}.
In comparison, the proposed Algorithm 1 completely avoids such approximation
and would be more stable in practical applications, as demonstrated by the results in Fig. 2, 3 and Table II in the experiments.

\section{Experimental Results}

We firstly investigate the robustness of the new algorithm
in terms of sensitivity against the kernel size parameter,
in comparison with the most close method \cite{13}.
Then, we evaluate the proposed algorithm on three benchmark datasets
in comparison with state-of-the-art blind image deblurring methods.
Furthermore, we conduct evaluation on face, natural, text, and low-light images.
Matlab code for reproducing the results of the new algorithm is available at \textit{https://github.com/FWen/deblur-pmp.git}.

For the new algorithm, $\mu = 4 \times {10^{-3}}$, $a = 2$, $J = 3$,
${\beta _0} = 2\mu $, and ${\beta _{max}} = {10^5}$ are used.
For each scale, we use $max\_iter=5$ as a trade-off between accuracy and speed.
The threshold parameter for PMP is initially set to $\lambda = 0.1$ and
gradually reduced to the mean of the PMP values.
Note that, in general the optimal values of these parameters depend on the distribution of the clear images,
the kernel models, the optimization algorithm used for the non-convex deblurring problem.
Hence, it is difficult to select the optimal values of these parameters.
Like most blind deblurring methods, such as the ones compared in the sequel, these parameters are selected empirically in practice.

The ``first loose and then tight'' strategy introduced in Section III-C
is employed to make the algorithm performing practically well.
The patch size is set dependent on the image size as $r = 0.025 \cdot mean(m,n)$.
Similar to \cite{1, 2, 5, 13}, we first estimate the blur kernel
by the proposed algorithm, and then obtain the final latent image
based on the estimated kernel by a non-blind deblurring method.
The non-blind deblurring algorithm \cite{14} is employed for
the final latent image estimation. The performance of the
compared algorithms is evaluated in terms of
peak-signal-to-noise ratio (PSNR),  structural similarity (SSIM) \cite{49} and cumulative error ratio
of the deblurred images and kernel estimation.

Table I presents a quantitative evaluation of the proposed method versus the patch size $r$ on the dataset \cite{12}.
It can be seen that the new method is robust to the patch size.

\begin{table}[!t]
\renewcommand\arraystretch{1.05}
\caption{Quantitative evaluation of the new method versus patch size on the dataset \cite{12} (average PSNR and SSIM).}
\centering
\begin{tabular}{|l|c|c|}
\hline
 Patch size &  PSNR (dB) & SSIM \\
\hline
$0.015\cdot mean(m,n)$ & 29.5891    & 0.8754      \\
\hline
$0.020\cdot  mean(m,n)$ & 29.6397    &   0.8778    \\
\hline
$0.025\cdot  mean(m,n)$ &	29.9764   &   0.8944  \\
\hline
$0.030\cdot  mean(m,n)$ &  29.7355    &    0.8837   \\
\hline
$0.035\cdot  mean(m,n)$ 	&  29.7088 &   0.8822 \\
\hline
\end{tabular}
\end{table}

\subsection{Robustness: Sensitivity to the Kernel Size Parameter}

As discussed in Section IV, the new algorithm
can avoid the non-rigorous approximation in solving non-explicit priors
(e.g., dark-channel) involved subproblems in [10], [12], [14].
This brings the new algorithm a potential advantage of being more stable in practical applications.
To illustrate this point, the first experiment compares the performance of the new algorithm with 
Pan \textit{et al.} [10] through investigating their sensitivity against the kernel size parameter.
The selection of the kernel size parameter has a substantial influence on the performance of most deblurring algorithms.

\begin{table}[!t]
\renewcommand\arraystretch{1.05}
\caption{Average PSNR and SSIM on deblurring 12 samples of the dataset \cite{12} associated with the kernels 8, 9 and 10.}
\centering
\begin{tabular}{|l|c|c|c|c|c|}
\hline
 Kernel size   &  111  &   121 &  131 & 141 & 151 \\
\hline
\hline
  Pan \textit{et al.} (PSNR) & 22.40  & 22.85 &  21.71 &  23.13 &  21.69\\
\hline
 Ours (PSNR) & 23.05  & 23.35  & 23.42 &  23.39  & 22.49\\
\hline
\hline
  Pan \textit{et al.} (SSIM) & 0.7503  &  0.7651  &  0.7087  &  0.7612  &  0.7141\\
\hline
 Ours (SSIM)& 0.7540  &  0.7708  &  0.7806  &  0.7619  &  0.7148\\
 \hline
\end{tabular}
\end{table}

\begin{figure}[!t]
\centering
\subfigure[PSNR]{\includegraphics[width=4.25cm]{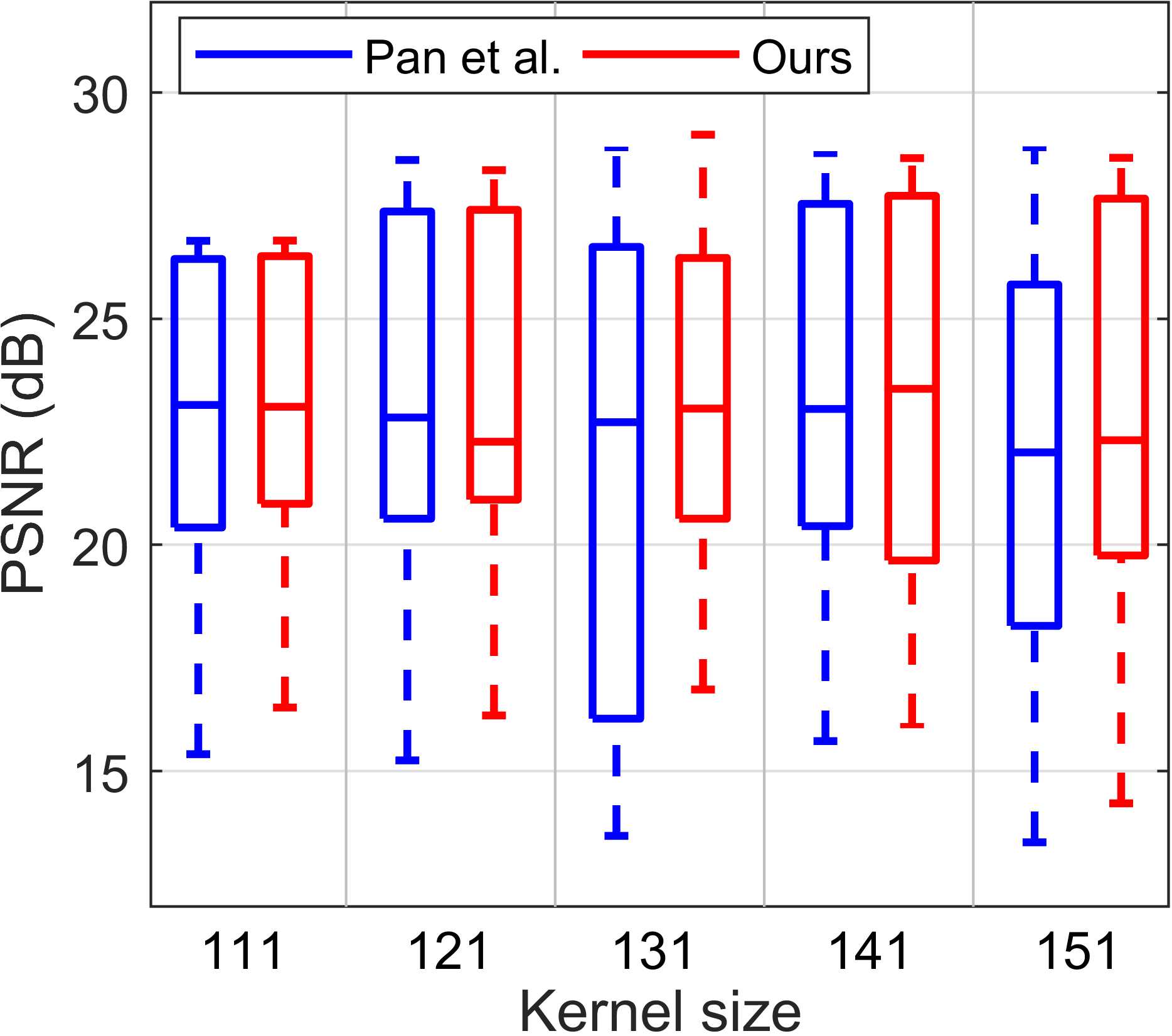}}~
\subfigure[SSIM]{\includegraphics[width=4.25cm]{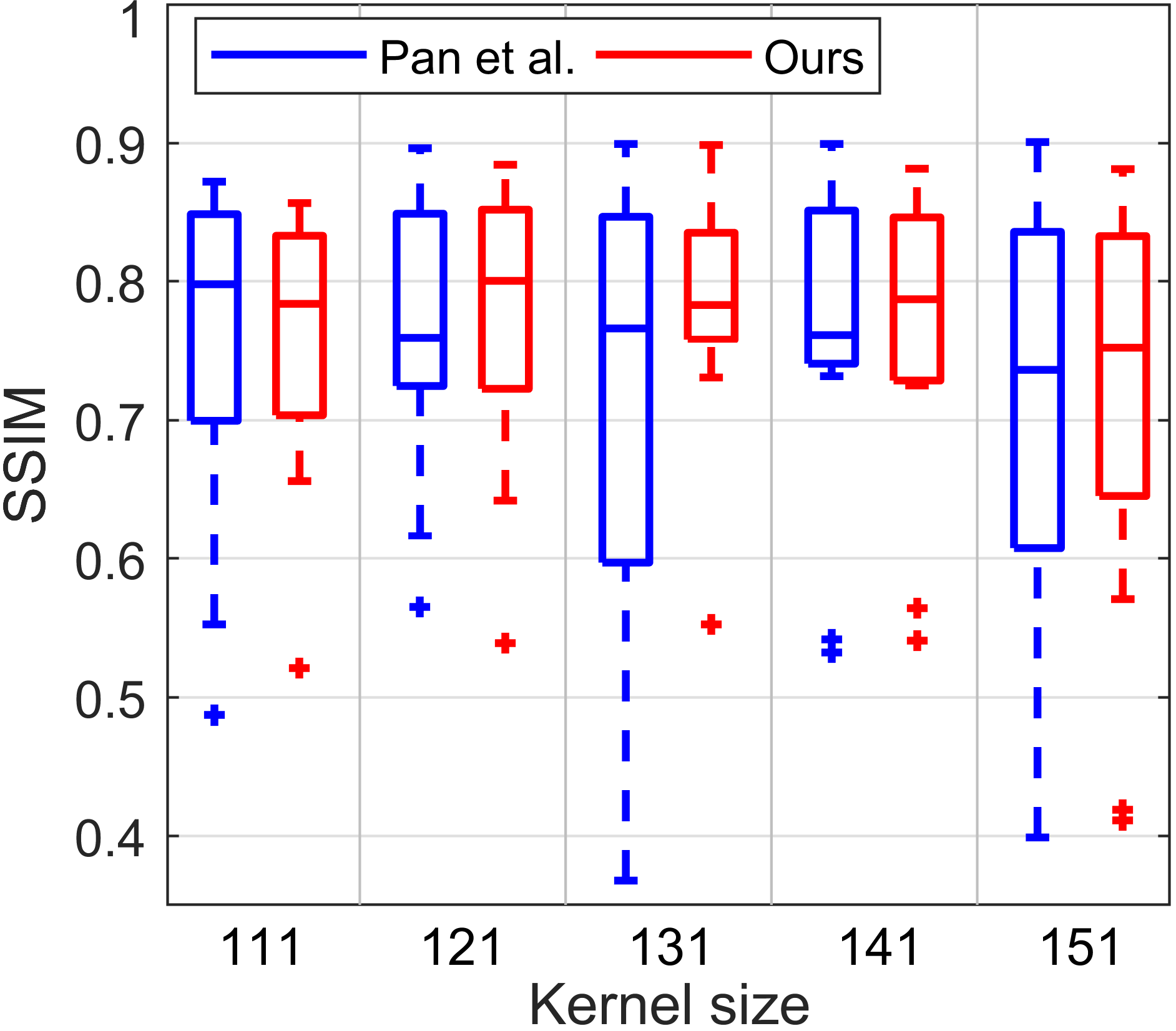}}
\caption{{PSNR and SSIM versus kernel size on deblurring 12 samples of the dataset \cite{12} associated with the kernels 8, 9 and 10.}}
\label{figure2_1}
\end{figure}


\begin{figure*}[!t]
\centering
\subfigure[The first image. From left to right, the used kernel sizes are \{25, 35, 45, 55, 65\}.]
{\includegraphics[width = 7in]{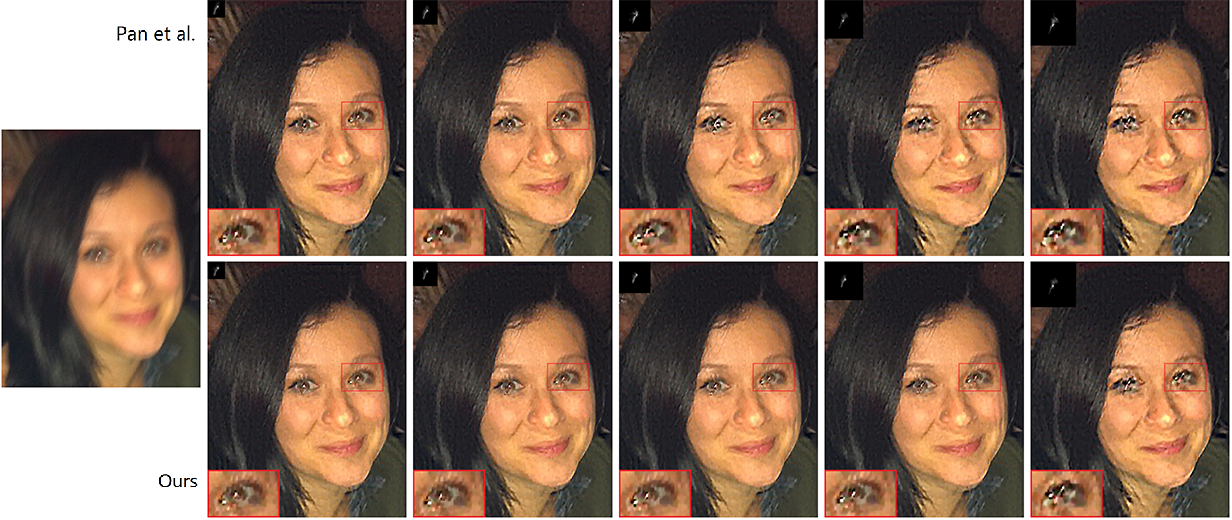}}\\
\subfigure[The second image. From left to right, the used kernel sizes are \{45, 55, 65, 75, 85\}.]
{\includegraphics[width = 7in]{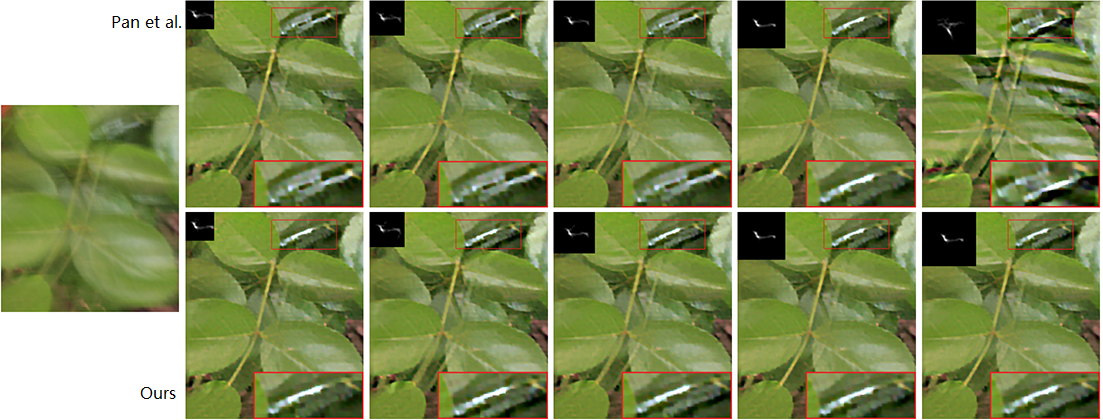}}\\
\subfigure[The third image. From left to right, the used kernel sizes are \{65, 75, 85, 95, 105\}.]
{\includegraphics[width = 7in]{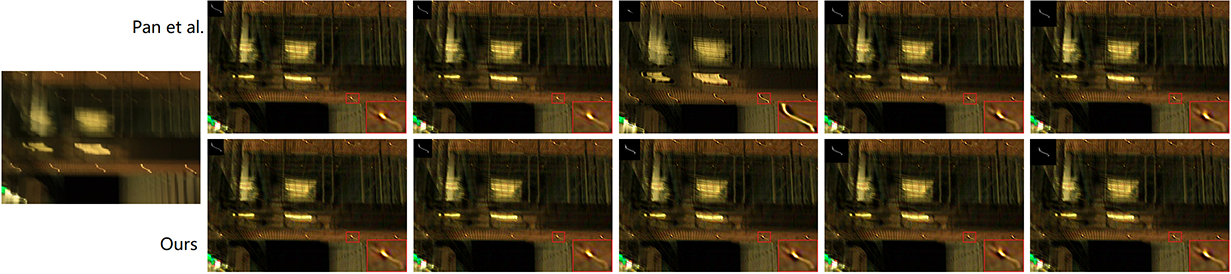}}
\caption{Comparison between Pan \textit{et al.} \cite{13} and ours method on the three blurred images. For each method, different kernel sizes are tested.}
\label{figure2}
\end{figure*}

Table II and Fig. 2 present the PSNR and SSIM results of the two compared methods versus kernel size
on deblurring three challenging kernels in the dataset \cite{12} (the kernels 8, 9 and 10).
These selected kernels are the most challenging kernels in the dataset \cite{12},
which have sizes of more than 100 pixels. There are four samples for each of the three kernels.
Moreover, Fig. \ref{figure2} shows deblurred results of the two algorithms
when using different values of the kernel size parameter on three realistic images.
It can be seen from Fig. 2, \ref{figure2}, and Table II that our method is less sensitive to the kernel size variation,
which demonstrates its better robustness in practice.

\subsection{Evaluation on Benchmark Datasets}

\begin{figure}[t]
\centering
~~~~{\includegraphics[width=8.2cm]{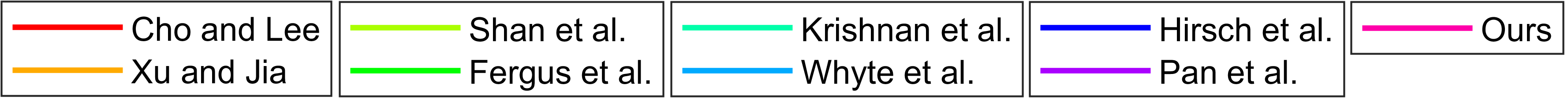}}\\\vskip 0.05cm
\subfigure[PSNR]{\includegraphics[width=4.25cm]{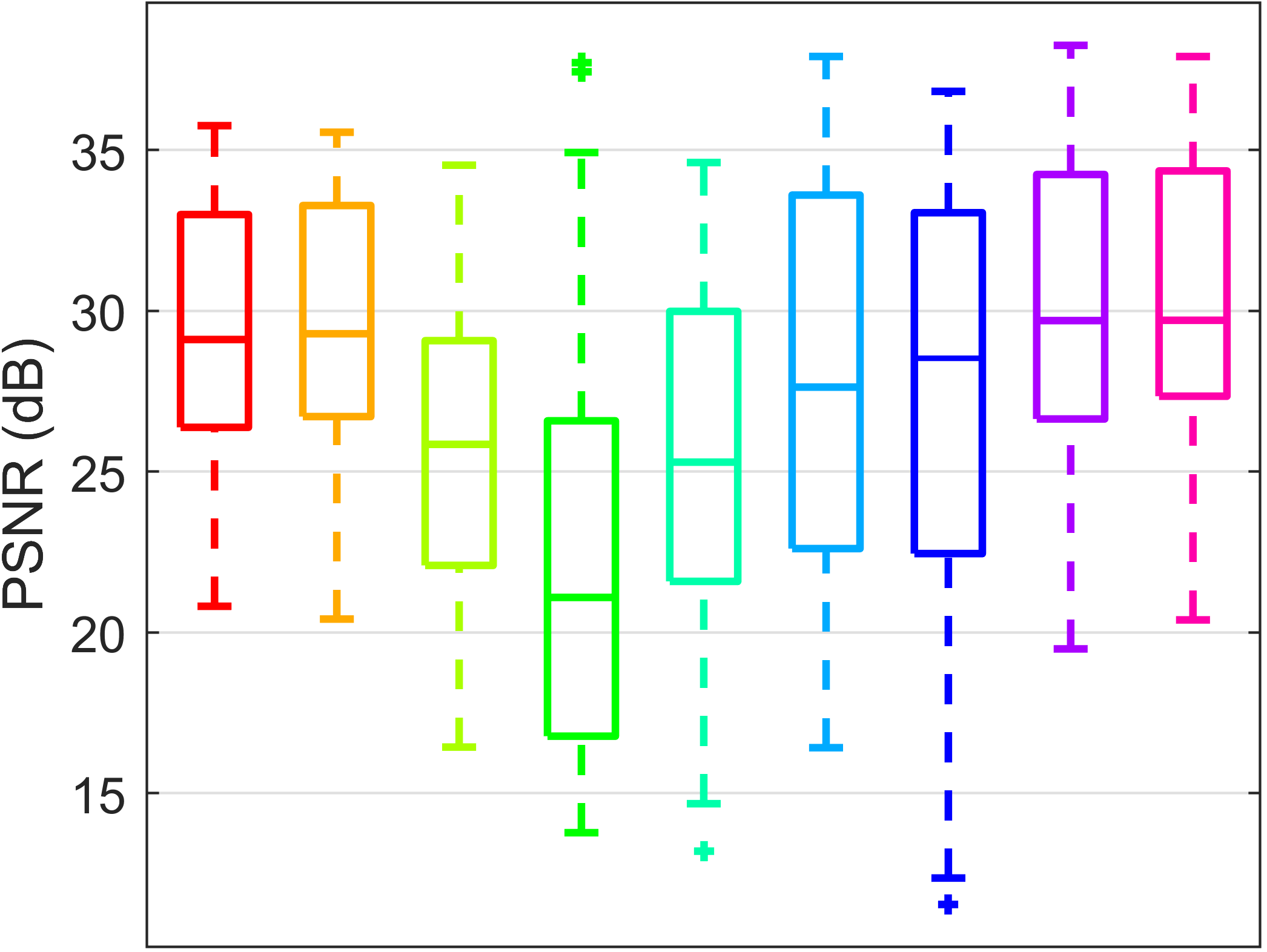}}~
\subfigure[SSIM]{\includegraphics[width=4.25cm]{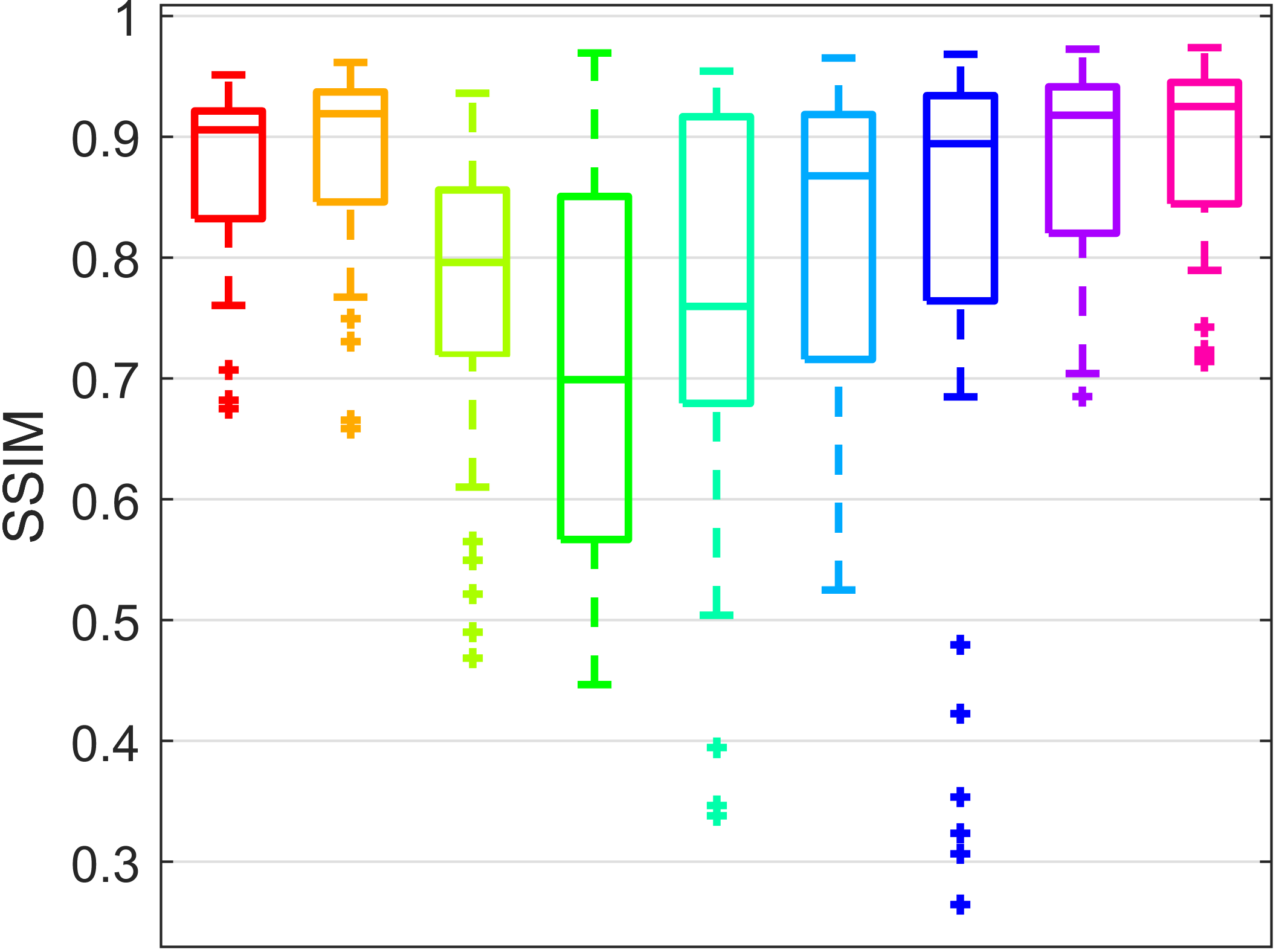}}
\caption{{Quantitative evaluation results on the benchmark dataset of Kohler \textit{et al.} \cite{12} (PSNR and SSIM comparison over 48 blurry images).}}
\label{figure3}
\end{figure}

The second experiment uses the dataset by Kohler \textit{et al.} \cite{12},
which contains 48 blurred samples corresponding to 4 clear images and 12 blur kernels.
The compared algorithms include Cho and Lee \cite{3}, Xu and Jia \cite{2},
Shan \textit{et al.} \cite{4}, Fergus \textit{et al.} \cite{1}, Krishnan \textit{et al.} \cite{6},
Whyte \textit{et al.} \cite{7}, Hirsch \textit{et al.} \cite{8}, and Pan \textit{et al.} \cite{13}.
Fig. \ref{figure3} presents a statistical analysis of the PSNR and SSIM results of the compared algorithms on deblurring the 48 blurred images.
Table III shows the average PSNR and average SSIM of the algorithms.
The PSNR and SSIM of each deblurred image are computed via comparing it with 199
clear images captured within the camera motion trajectory.
The results of the methods \cite{1,2,3,4, 6,7,8}
are those reported in \cite{12}, while the result of the method \cite{13}
is computed from the deblurred results provided by the authors at their
website\footnote{http://vllab1.ucmerced.edu/$\sim$jinshan/projects/dark-channel-deblur/}.

\begin{figure*}[t]
\centering
{{\includegraphics[width = 1.525in]{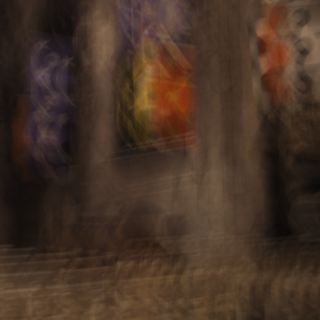}} {\includegraphics[width = 1.525in]{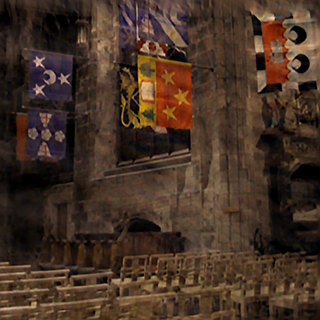}} {\includegraphics[width = 1.525in]{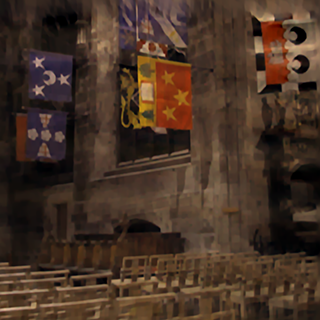}} {\includegraphics[width = 1.525in]{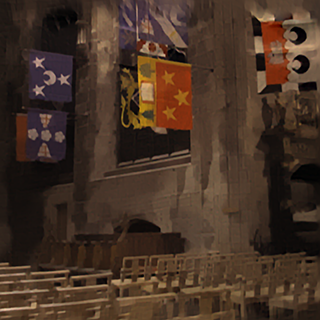}}}\\\vskip 2pt
{{\includegraphics[width = 1.525in]{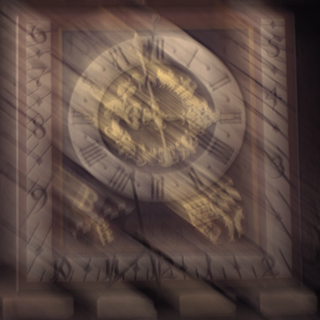}} {\includegraphics[width = 1.525in]{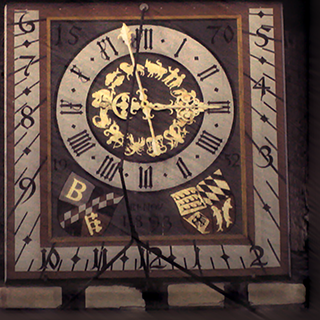}} {\includegraphics[width = 1.525in]{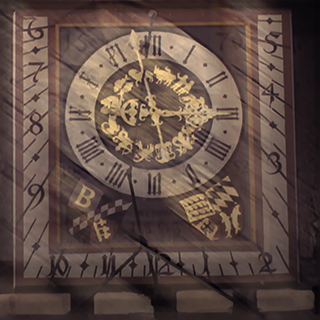}} {\includegraphics[width = 1.525in]{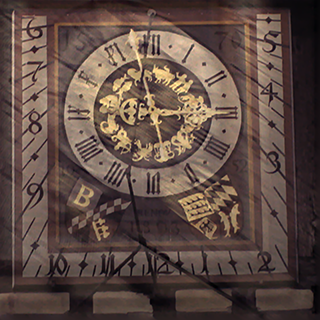}}}\\\vskip 2pt
{{\includegraphics[width = 1.525in]{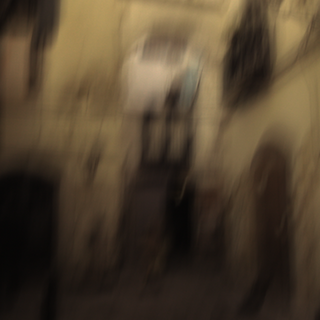}} {\includegraphics[width = 1.525in]{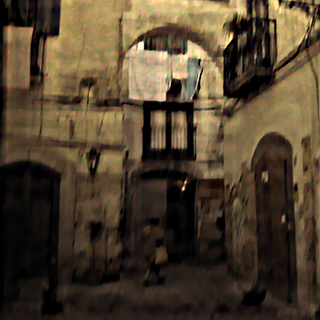}} {\includegraphics[width = 1.525in]{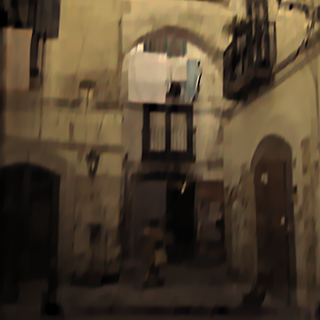}} {\includegraphics[width = 1.525in]{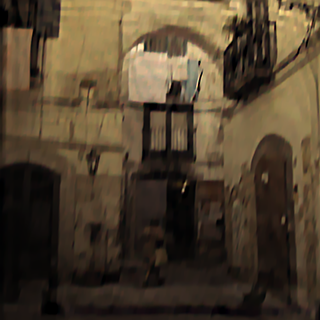}}}\\\vskip 2pt
\subfigure[Blurred image]{{\includegraphics[width = 1.525in]{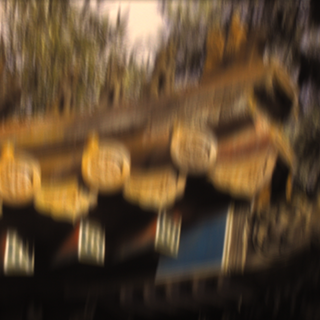}}}
\subfigure[Xu \textit{et al.} {\cite{9}}]{{\includegraphics[width = 1.525in]{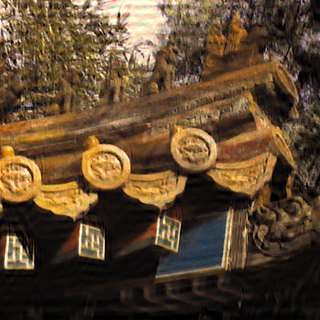}}}
\subfigure[Pan \textit{et al.} {\cite{13}}]{{\includegraphics[width = 1.525in]{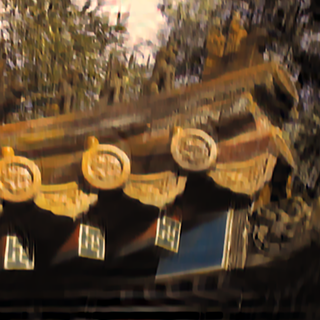}}}
\subfigure[Ours]{{\includegraphics[width = 1.525in]{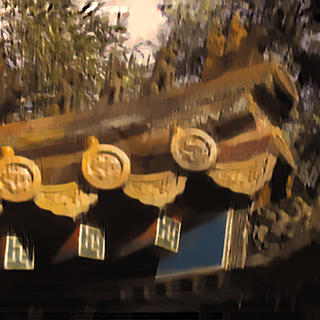}}}
\caption{Visual comparison on four challenging images from the dataset \cite{12}. From top to bottom are, respectively, the `Blurry1\_8', `Blurry2\_9', `Blurry3\_10', and `Blurry4\_11' images.}
\label{figure4}
\end{figure*}

It can be seen that the new algorithm can achieve state-of-the-art
performance in terms of the PSNR and SSIM results.
Fig. \ref{figure4} presents visual comparison
on four challenging images with heavy blurs from the dataset \cite{12},
including the `Blurry1\_8', `Blurry2\_9', `Blurry3\_10', and `Blurry4\_11' images.
It can be inferred from Fig. \ref{figure4} that the proposed algorithm can achieve comparable
or even better visual results compared with existing state-of-the-art methods \cite{9,13}.

\begin{table}[!t]
\renewcommand\arraystretch{1.05}
\caption{Quantitative results on the dataset of Kohler \textit{et al.} \cite{12}, including the average PSNR and average SSIM.}
\centering
\begin{tabular}{|l|c|c|}
\hline
 Method &  PSNR (dB) & SSIM \\
\hline
Cho and Lee & 28.9831 & 0.8746  \\
\hline
Xu and Jia & 29.5373 &   0.8851   \\
\hline
Shan \textit{et al.}& 25.8912 &  0.7748   \\
\hline
Fergus \textit{et al.}&	22.7303 &  0.7048    \\
\hline
Krishnan \textit{et al.} & 25.7246 & 0.7608\\
\hline
Whyte \textit{et al.}	&  27.8441 & 0.8116\\
\hline
Hirsch \textit{et al.}	& 26.8388  & 0.8095\\
\hline
Pan \textit{et al.}	& 29.9513 & 0.8853\\
\hline
Ours & \textbf{29.9764} & \textbf{0.8944}\\
\hline
\end{tabular}
\end{table}

\begin{figure}[!t]
\centering
{\includegraphics[width=2.9cm]{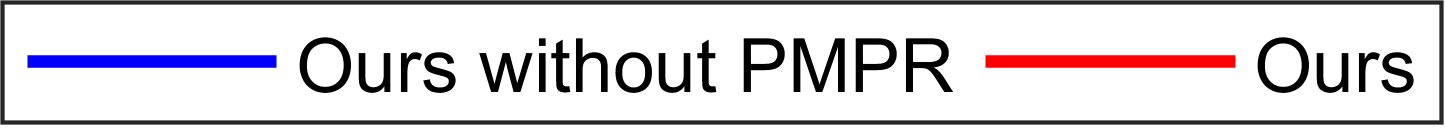}}~~~~~~~~~~~~~~~~~~~~~~~~~~~~~~~~~~~~~~~~~\\\vskip -0.01pt
\subfigure[]
{\includegraphics[width=2.95cm]{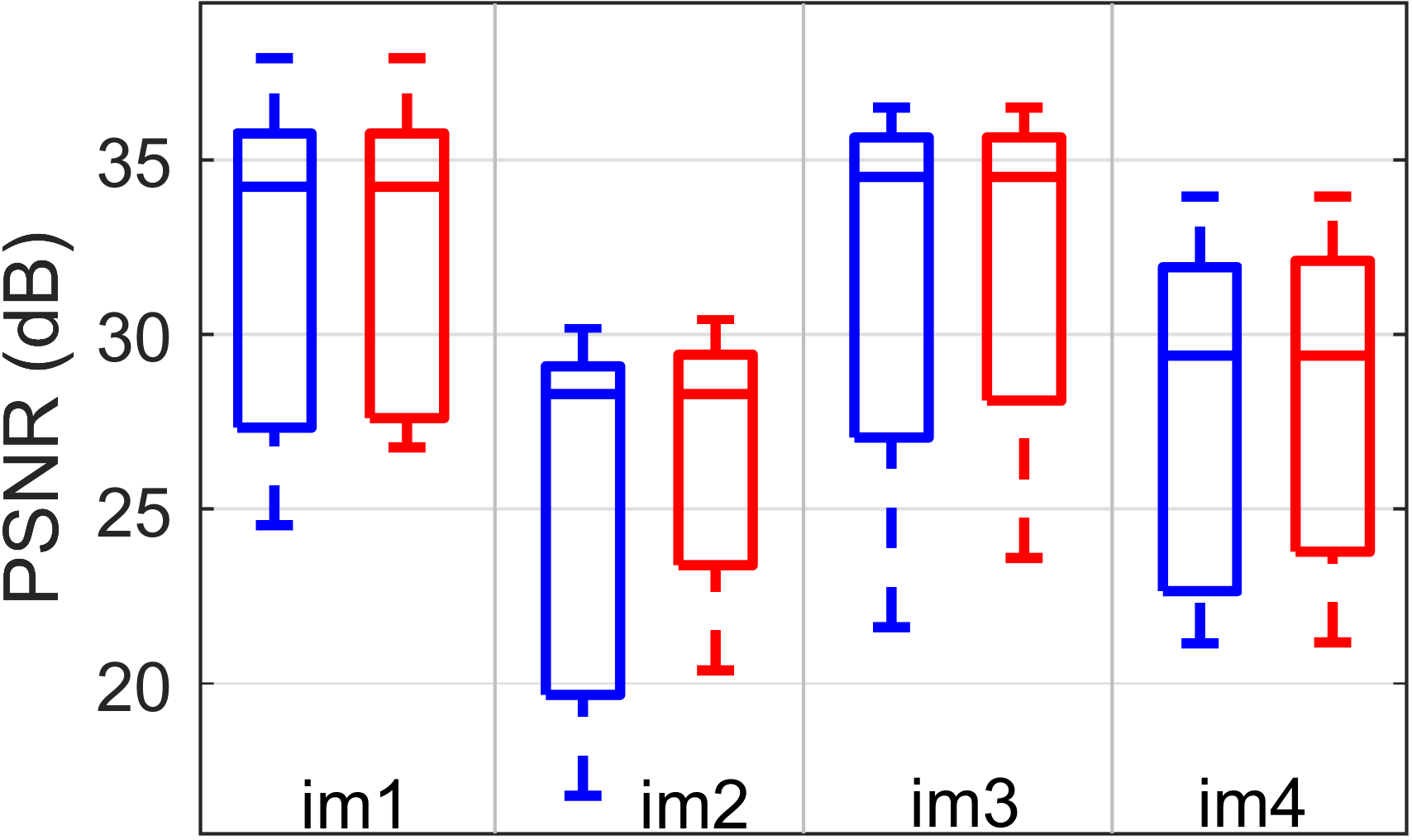}}~
\subfigure[]
{\includegraphics[width=2.95cm]{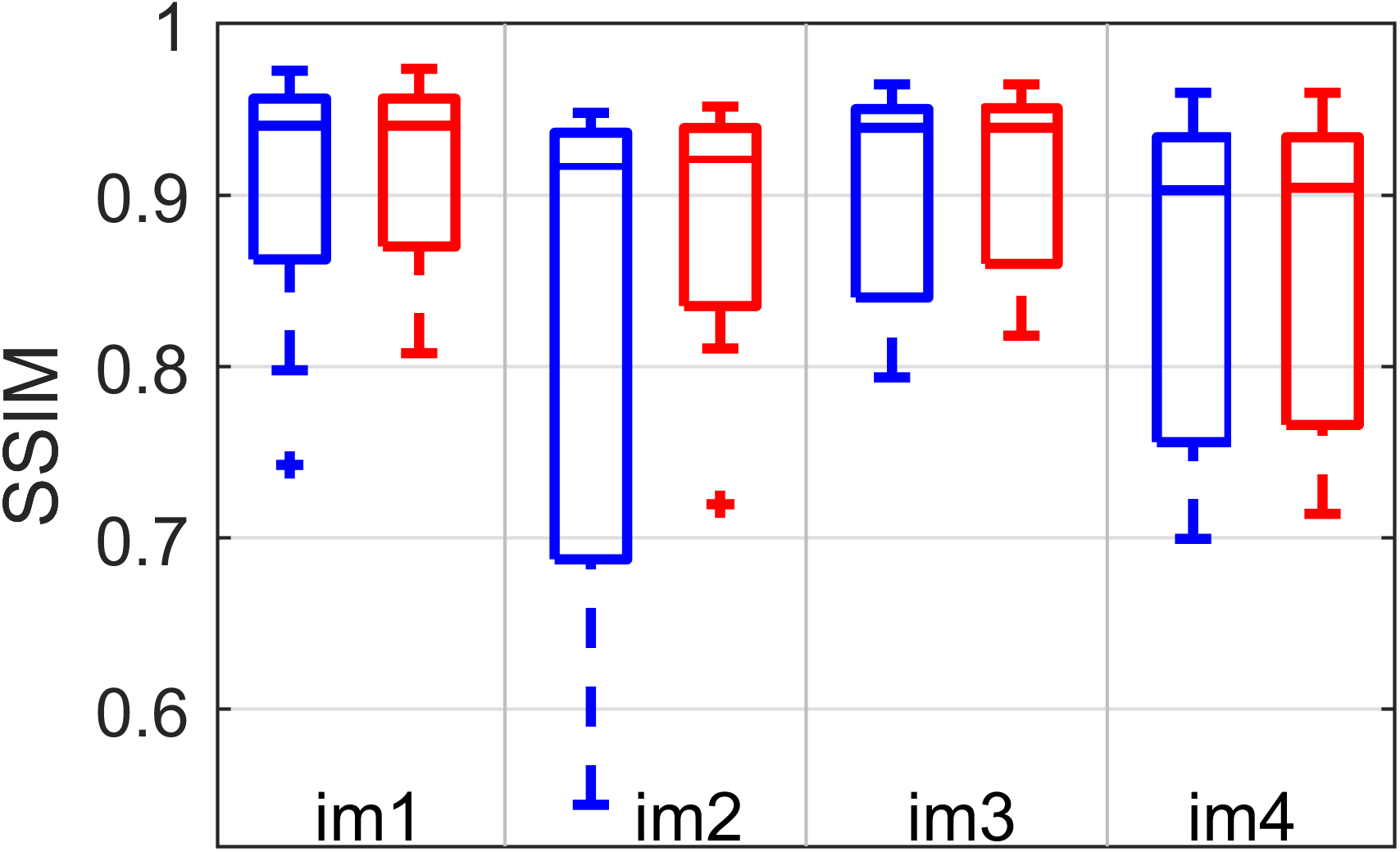}}~
\subfigure[]
{\includegraphics[width=1.1cm]{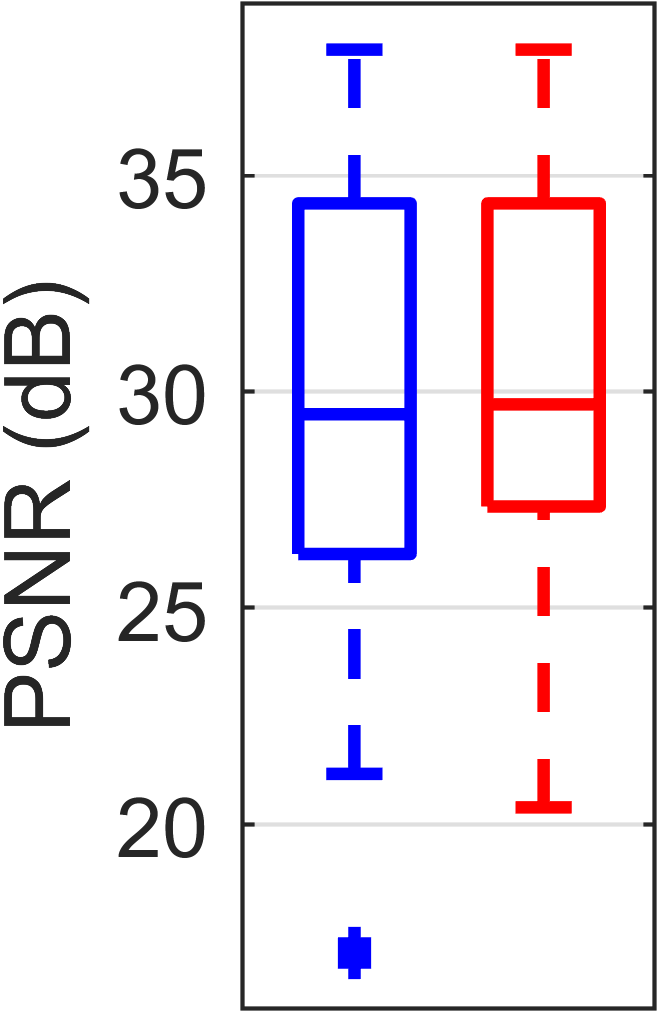}}~
\subfigure[]
{\includegraphics[width=1.1cm]{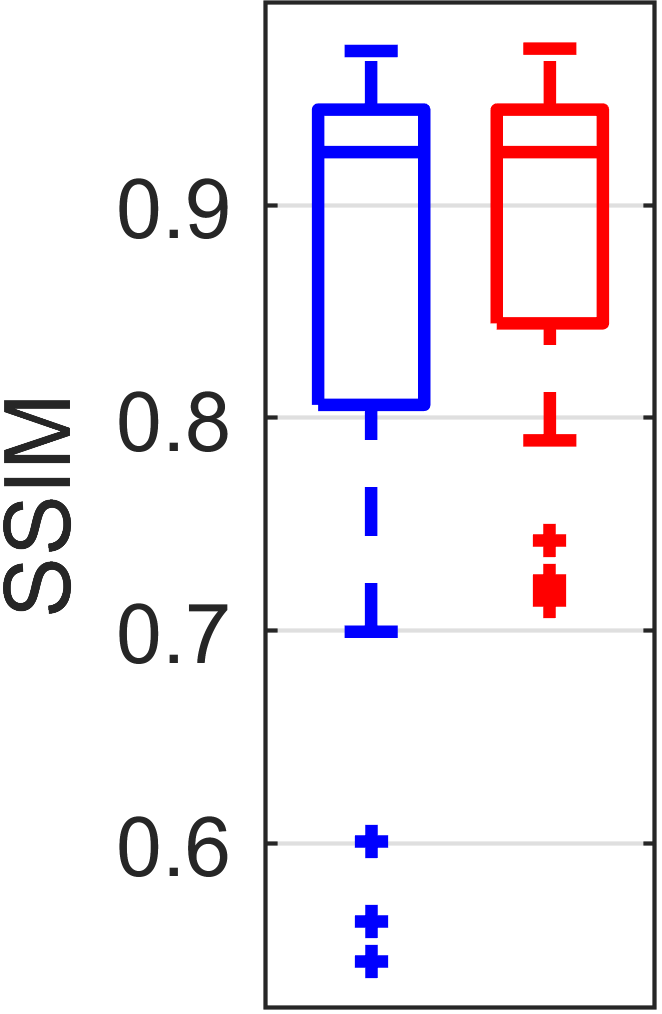}}
\caption{Quantitative results of the proposed algorithm with or without PMP regularization on the dataset \cite{12}. (a), (b): PSNR and SSIM over 12 blurry samples of each of the 4 images. (c), (d): PSNR and SSIM over all the 48 blurry samples.}
\label{figure5}
\end{figure}

Fig. \ref{figure5} further investigates the effectiveness of the proposed PMP
regularization to show the results of the new algorithm with and
without PMP regularization on the benchmark dataset \cite{12}.
The results demonstrate that the PMP regularization gives rise
to distinct PSNR and SSIM improvement.

\begin{figure}[!t]
\centering
{\includegraphics[width=7cm]{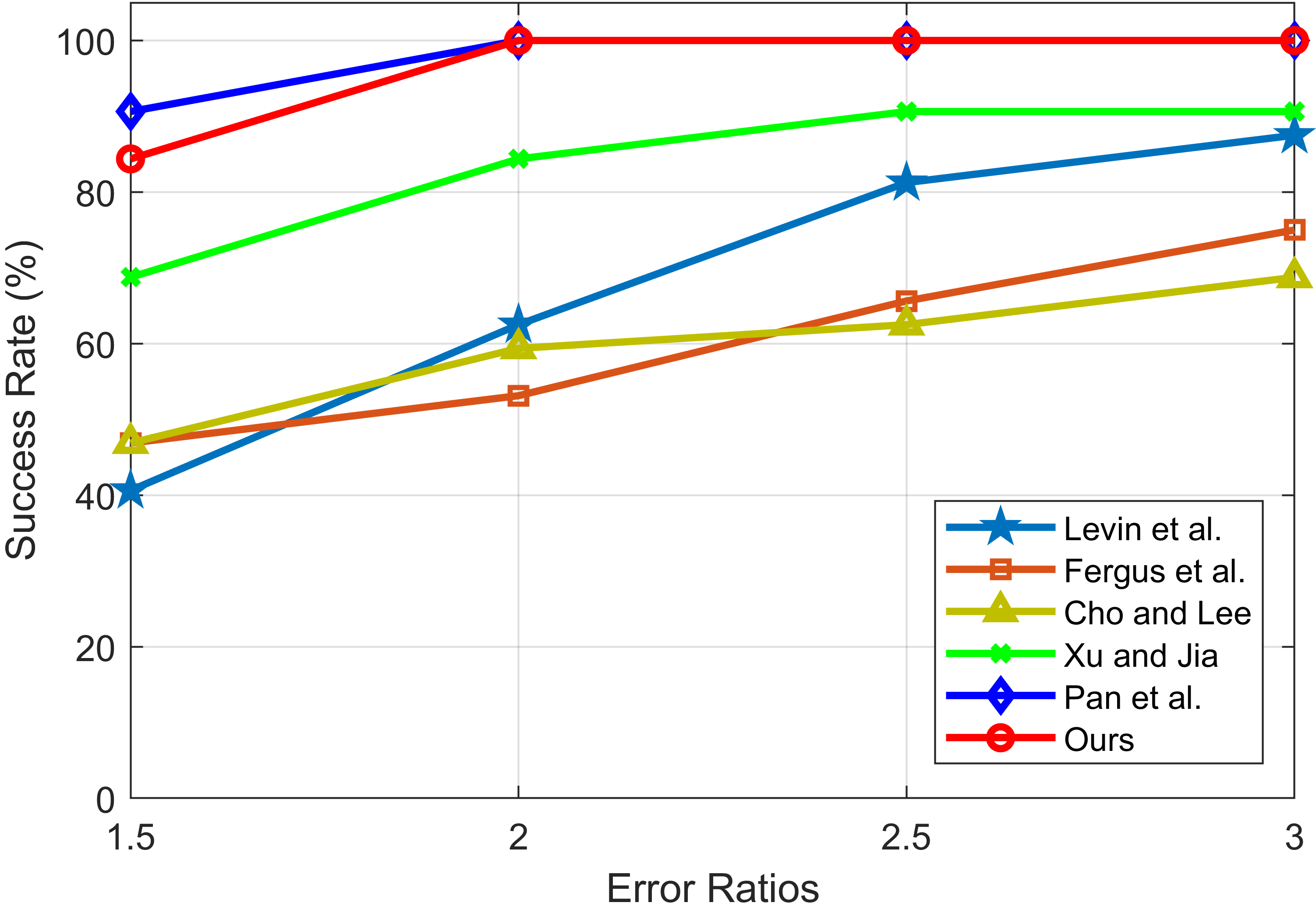}}
\caption{{Cumulative error ratios of the compared algorithms on the dataset of Levin \textit{et al.} \cite{5}.}}
\label{figure6}
\end{figure}

\begin{figure}[!t]
\centering
\subfigure[PSNR]
{\includegraphics[width=4.2cm]{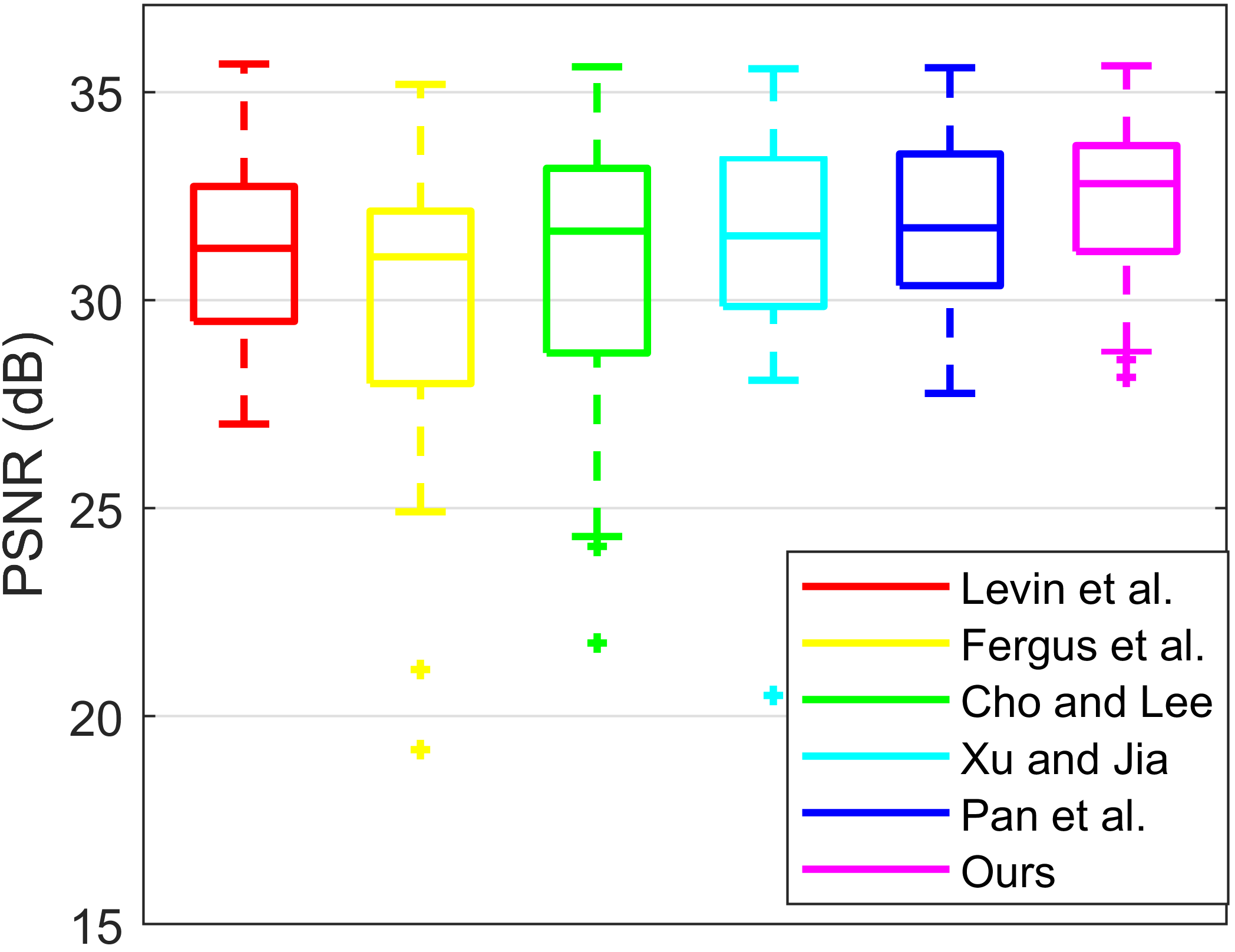}}~
\subfigure[SSIM]
{\includegraphics[width=4.2cm]{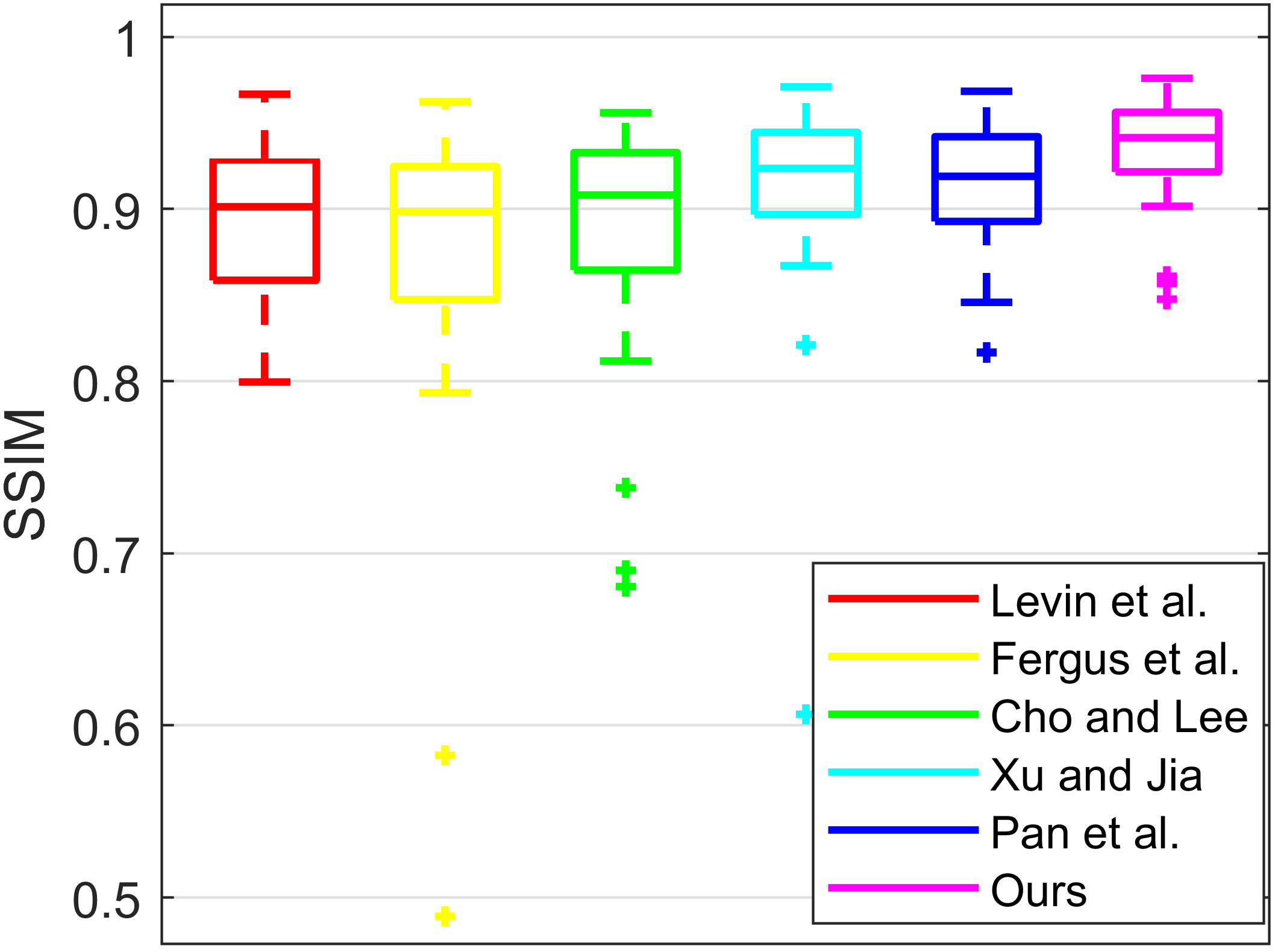}}
\caption{{Quantitative evaluation results on the benchmark dataset of Levin \textit{et al.} \cite{5} (PSNR and SSIM comparison over 32 blurry images).}}
\label{figure7}
\end{figure}

The third experiment uses the dataset by Levin \textit{et al.} \cite{5},
which contains 32 blurred samples corresponding to 4 clear images and 8 blur kernels.
The parameter $\mu$ is set to $5 \times {10^{ - 3}}$ for all examples.
Fig. \ref{figure6} shows the cumulative error ratios of the compared algorithms,
which are computed based on the sum of square difference (SSD) error as follows.
Firstly, for a restored image, the SSD error is computed as the SSD between it and its clear counterpart using the best shift between them.
Then, this SSD error is normalized with respect to the SSD error of the de-convolution result using the ground-truth kernel, which results in an error ratio.
Finally, from the error ratios of all the deblurred images, the cumulative error ratio is computed to get the success rate.
It is empirically noticed that deblurred results with
error ratios above 2 are visually implausible.
Table IV compares the average PSNR and average SSIM of the methods, whilst Fig. \ref{figure7} presents statistical analysis of the PSNR and SSIM results

\begin{table}[!t]
\renewcommand\arraystretch{1.05}
\caption{Quantitative results on the dataset of Levin \textit{et al.} \cite{5}, including the average PSNR and average SSIM.}
\centering
\begin{tabular}{|l|c|c|}
\hline
 Method &  PSNR (dB)  &  SSIM \\
\hline
Levin \textit{et al.} & 31.1372 & 0.8960\\
\hline
Fergus \textit{et al.} & 29.4629 & 0.8451\\
\hline
Cho and Lee & 30.7927 & 0.8837\\
\hline
Xu and Jia & 31.3604 & 0.9083\\
\hline
Pan \textit{et al.} & 31.7297 & 0.9148\\
\hline
Ours & \textbf{32.4450} & \textbf{0.9344}\\
\hline
\end{tabular}
\end{table}

\begin{figure}[!t]
\centering
{\includegraphics[width=8.0cm]{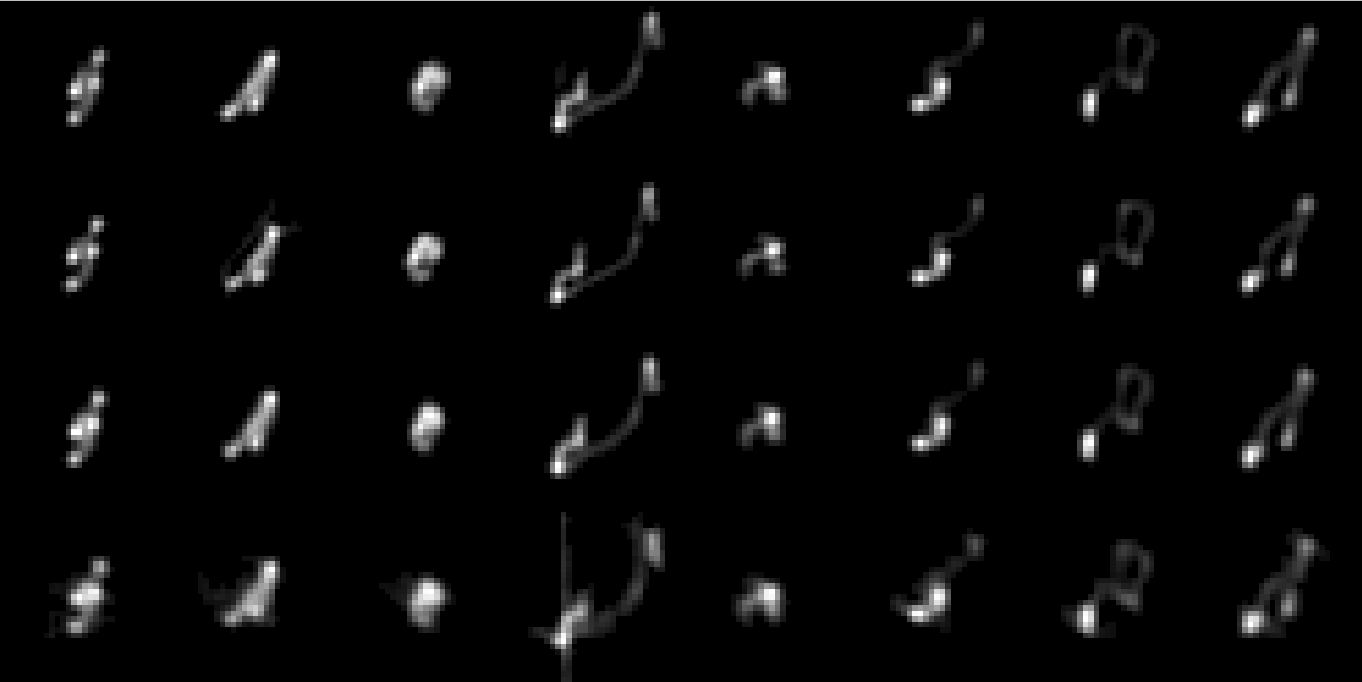}}
\caption{Estimated kernels by the proposed algorithm on the dataset \cite{5}.}
\label{figure8}
\end{figure}

\begin{figure}[!t]
\centering
\subfigure[im01\_ker04]{{\includegraphics[width = 1.5in]{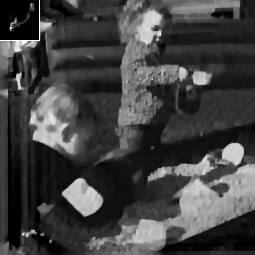}}}
\subfigure[im01\_ker06]{{\includegraphics[width = 1.5in]{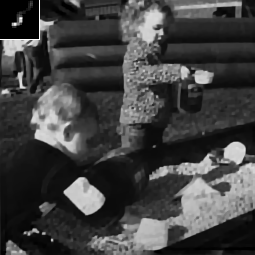}}}
\subfigure[im01\_ker07]{{\includegraphics[width = 1.5in]{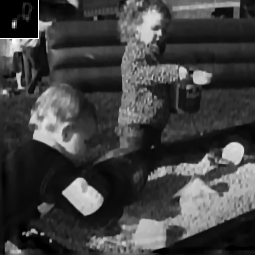}}}
\subfigure[im01\_ker08]{{\includegraphics[width = 1.5in]{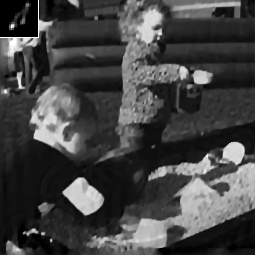}}}
\caption{Deblurred results by the proposed algorithm on four challenging samples in the dataset \cite{5}.}
\label{figure9}
\end{figure}

\begin{figure}[!t]
\centering
{\includegraphics[width=7cm]{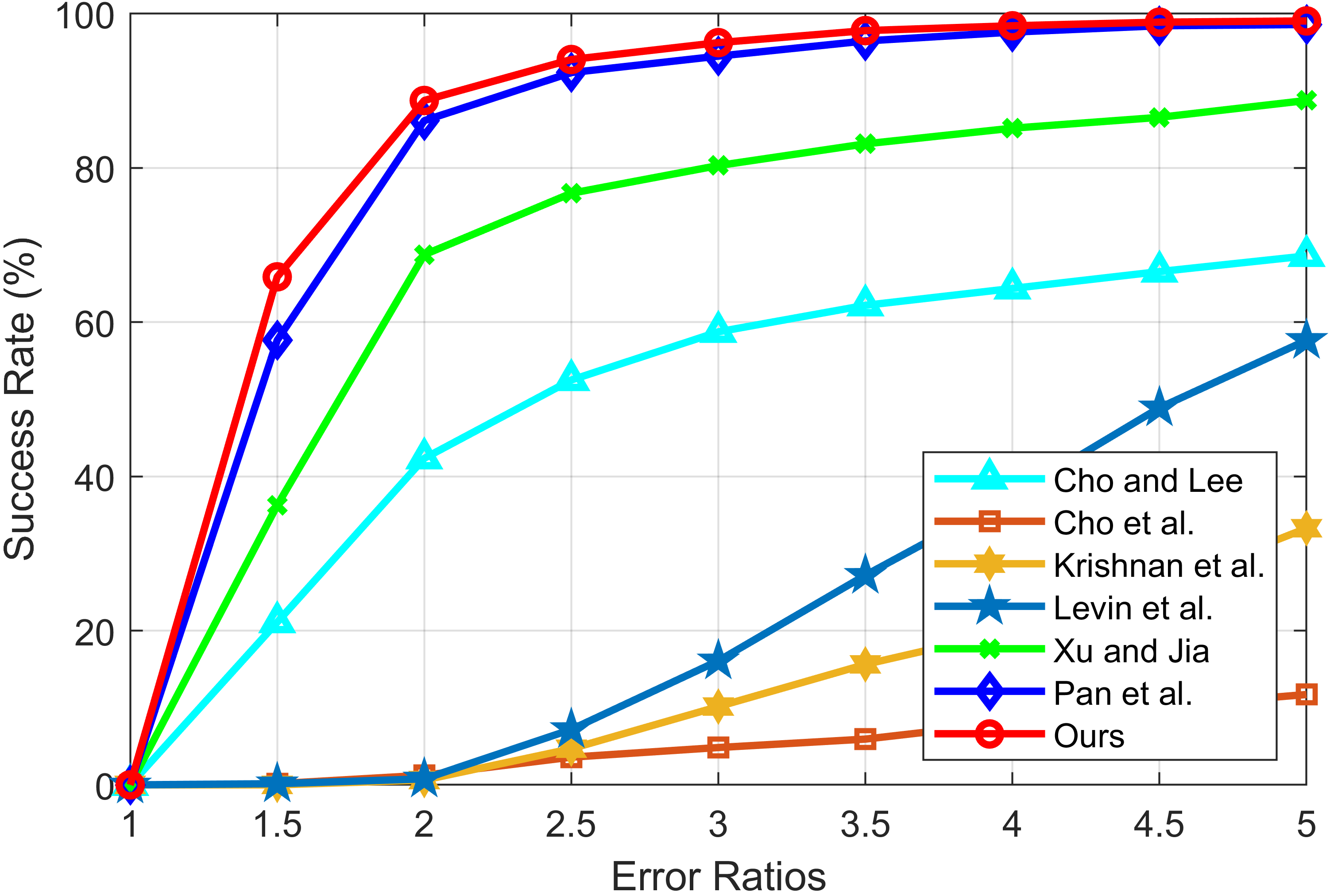}}
\caption{{Cumulative error ratios of the compared algorithms on the dataset \cite{11}.}}
\label{figure10}
\end{figure}

\begin{figure}[!t]
\centering
{\includegraphics[width=7.5cm]{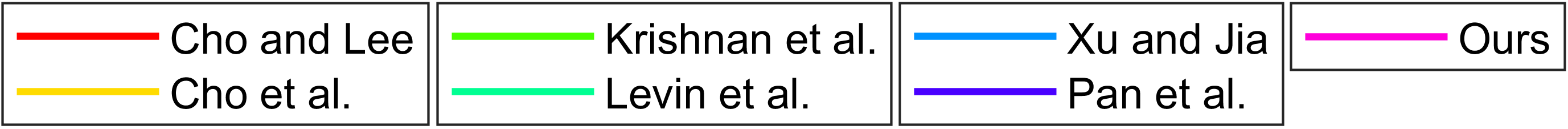}}\\\vskip 0.05cm
\subfigure[PSNR]{\includegraphics[width=4.25cm]{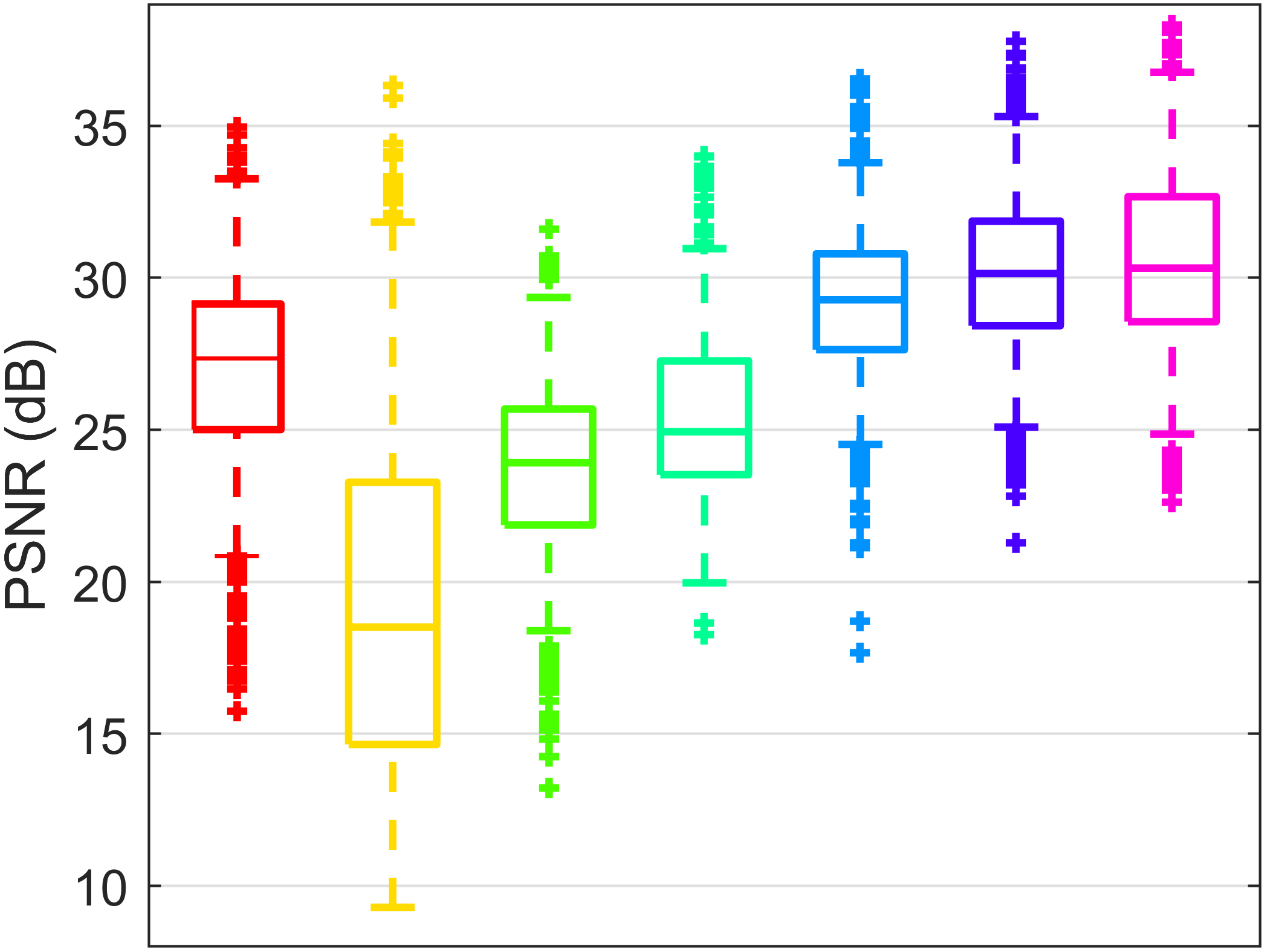}}~
\subfigure[SSIM]{\includegraphics[width=4.25cm]{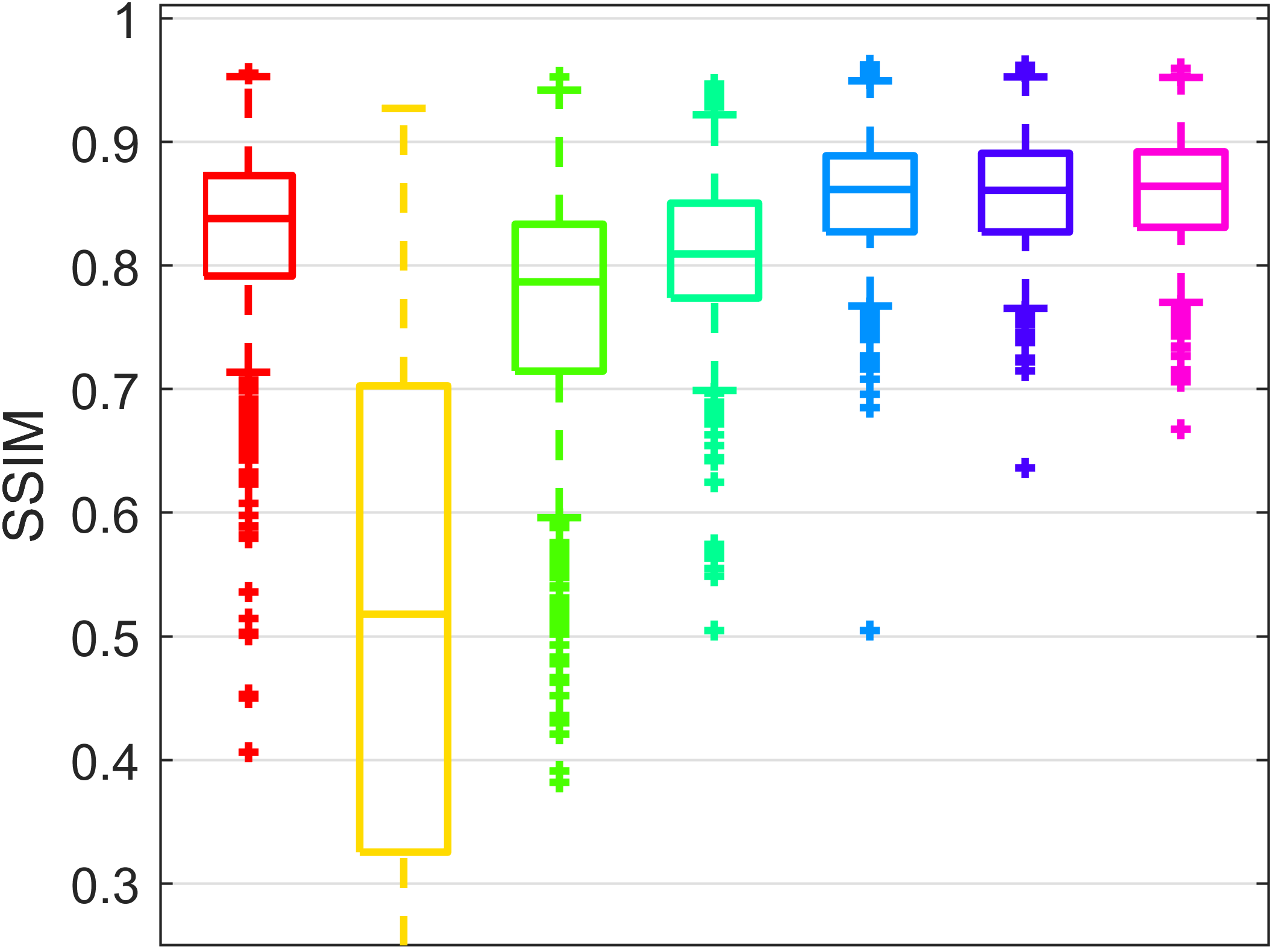}}
\caption{{Quantitative evaluation results on the benchmark dataset \cite{11} (PSNR and SSIM comparison over 640 blurry images).}}
\label{figure11}
\end{figure}

It can be observed from Table IV, Fig. \ref{figure6} and Fig. \ref{figure7} that,
the proposed algorithm achieves the best PSNR and SSIM results while attaining 100\% success at an error ratio of 2
on the dataset \cite{5}.
Fig. \ref{figure8} illustrates the estimated kernels by the
proposed algorithm on this dataset, whilst Fig. \ref{figure9} presents
the deblurred results of the proposed algorithm on
four challenging samples in this dataset.

The fourth experiment further considers a much larger dataset of Sun \textit{et al.} \cite{11},
which contains 640 blurred samples corresponding to 80 clear images and 8 blur kernels.
Fig. \ref{figure10} shows the cumulative error ratios of the algorithms on this dataset.
The compared algorithms include Cho and Lee \cite{3}, Cho \textit{et al.} \cite{35}, Krishnan \textit{et al.} \cite{6}, Levin \textit{et al.} \cite{5}, Xu and Jia \cite{2},
and Pan \textit{et al.} \cite{13}.
Table V compares the average PSNR and average SSIM of the algorithms, whilst Fig. \ref{figure11} presents statistical analysis of the PSNR and SSIM results.
It can be seen that the proposed algorithm can achieve the state-of-the-art performance in terms of cumulative error ratio, PSNR and SSIM of revcovery.

\subsection{Evaluation on Natural and Specific Images}

In what follows, the proposed method is further evaluated on face,
natural, text, and low-light images.
Here we only provide some typical results for each class
due to limited space. More samples are provided online at
\textit{https://github.com/FWen/deblur-pmp.git}.

\begin{figure*}[!t]
\centering
{{\includegraphics[width = 1.32in]{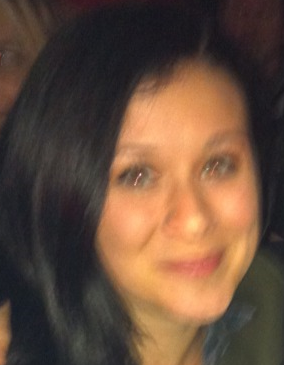}} {\includegraphics[width = 1.32in]{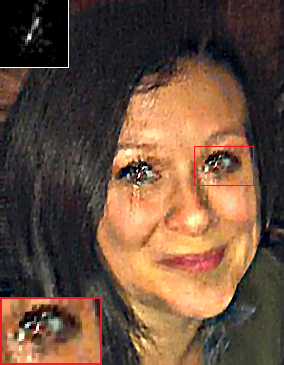}}
{\includegraphics[width = 1.32in]{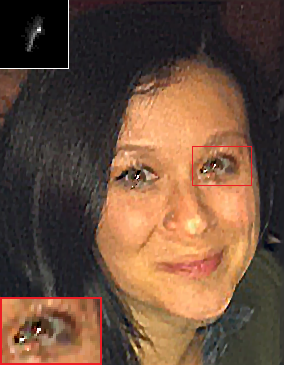}} {\includegraphics[width = 1.32in]{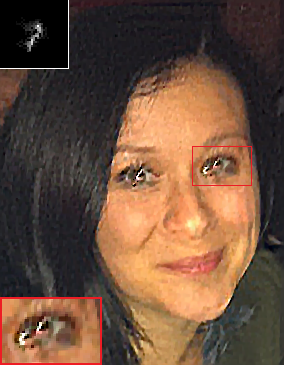}}
{\includegraphics[width = 1.32in]{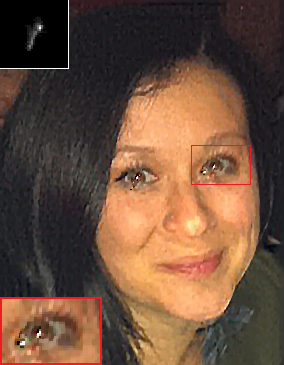}}}\\\vskip 3pt
\subfigure[Blurred image]{{\includegraphics[width = 1.32in]{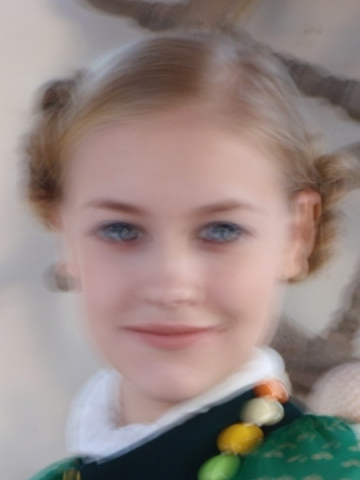}}}
\subfigure[Xu \textit{et al.} {\cite{9}}]{{\includegraphics[width = 1.32in]{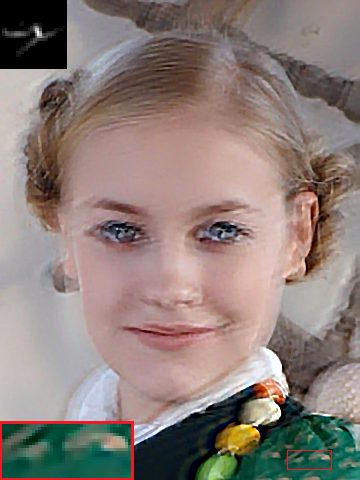}}}
\subfigure[Pan \textit{et al.} {\cite{13}}]{{\includegraphics[width = 1.32in]{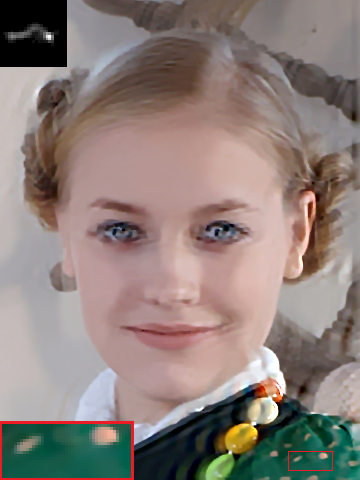}}}
\subfigure[Ours without PMP]{{\includegraphics[width = 1.32in]{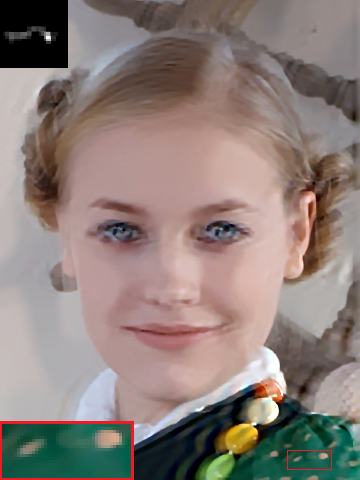}}}
\subfigure[Ours]{{\includegraphics[width = 1.32in]{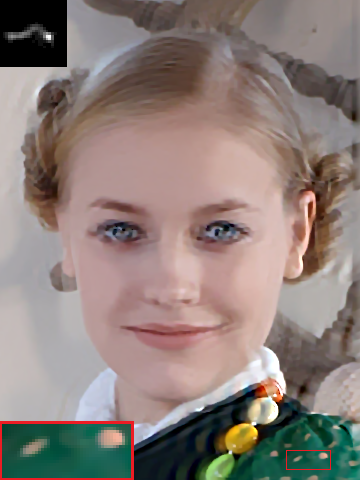}}}
\caption{Visual comparison on two realistic blurred face images.}
\label{figure12}
\end{figure*}

\textbf{Face image:}
Face image deblurring is challenging for methods developed for natural images,
since the lack of edges and textures in face images makes accurate
kernel estimation challenging. Fig. \ref{figure12} compares the proposed method
with the methods \cite{9, 13} on two realistic blurred face images.
The results demonstrate that our method compares favorably or even better against the methods give in \cite{9, 13}.

\begin{figure*}[!t]
\centering
{{\includegraphics[width = 1.32in]{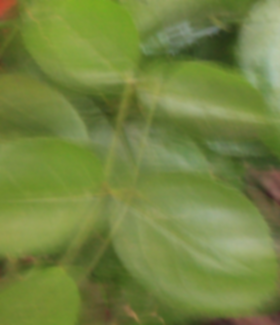}} {\includegraphics[width = 1.32in]{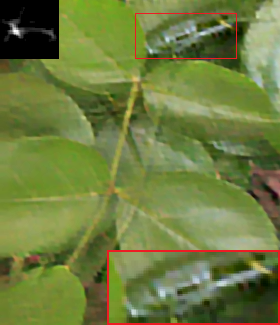}}
{\includegraphics[width = 1.32in]{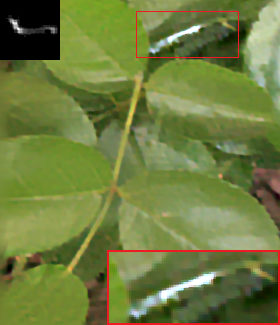}} {\includegraphics[width = 1.32in]{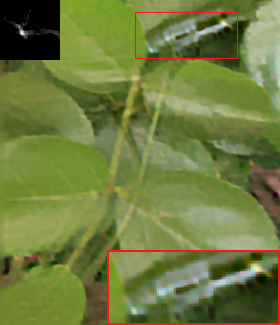}}
{\includegraphics[width = 1.32in]{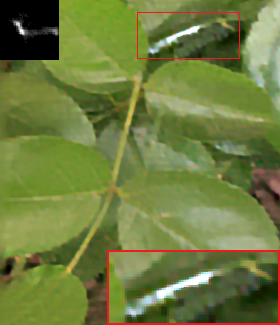}}}\\\vskip 3pt
\subfigure[Blurred image]{{\includegraphics[width = 1.32in]{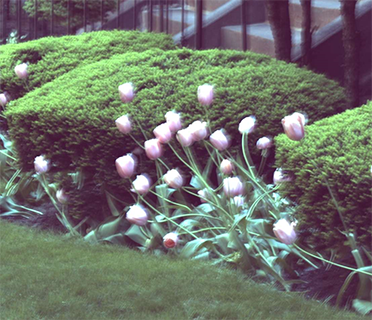}}}
\subfigure[Xu \textit{et al.} {\cite{9}}]{{\includegraphics[width = 1.32in]{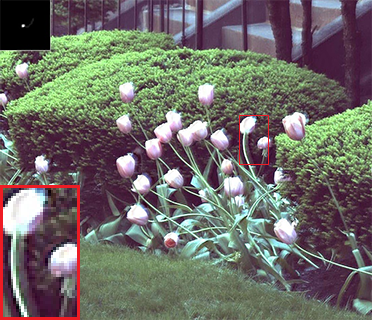}}}
\subfigure[Pan \textit{et al.} {\cite{13}}]{{\includegraphics[width = 1.32in]{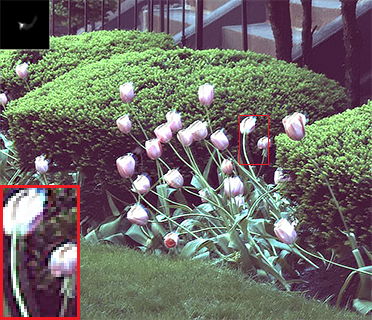}}}
\subfigure[Ours without PMP]{{\includegraphics[width = 1.32in]{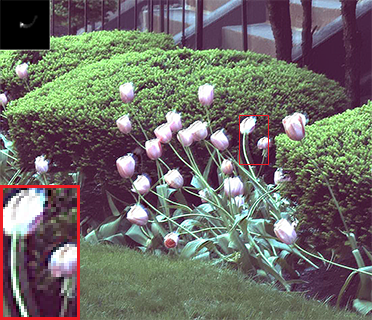}}}
\subfigure[Ours]{{\includegraphics[width = 1.32in]{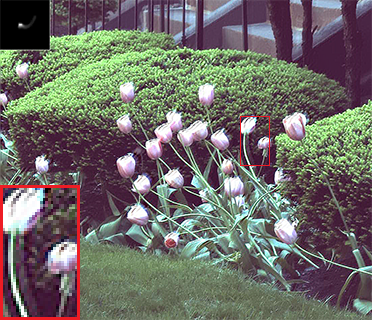}}}
\caption{Visual comparison on two real natural images.}
\label{figure13}
\end{figure*}

\textbf{Natural image:}
The results of the compared algorithms
on two real natural images are shown in Fig. \ref{figure13}.
Again, our algorithm compares competitively
against the methods \cite{9,13}. It can be see that
the proposed PMP regularization helps to significantly
reduce the ringing artifacts in the deblurred image,
which makes the proposed algorithm yielding state-of-the-art performance.

\begin{figure*}[!t]
\centering
{{\includegraphics[width = 1.32in]{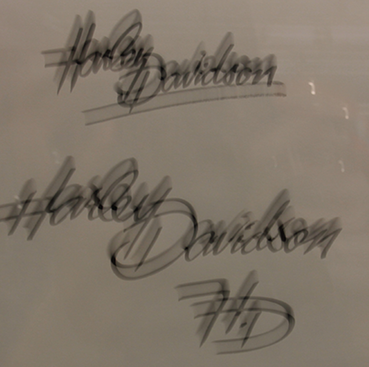}} {\includegraphics[width = 1.32in]{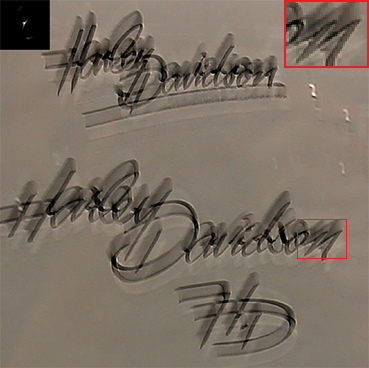}}
{\includegraphics[width = 1.32in]{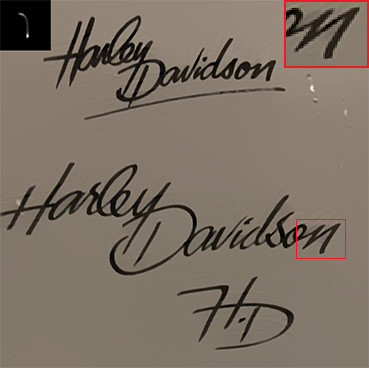}} {\includegraphics[width = 1.32in]{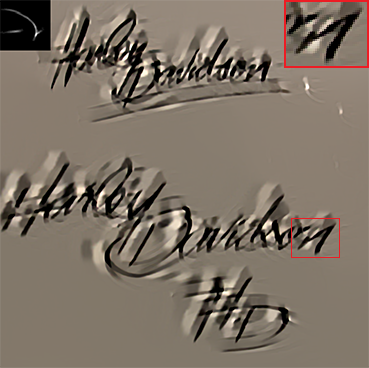}}
{\includegraphics[width = 1.32in]{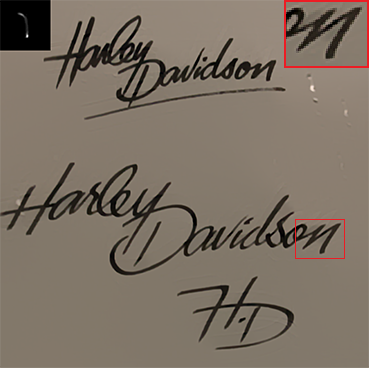}}}\\\vskip 3pt
\subfigure[Blurred image]{{\includegraphics[width = 1.32in]{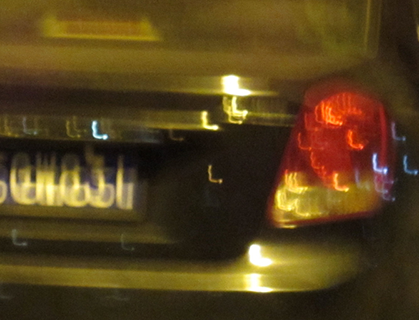}}}
\subfigure[Xu \textit{et al.} {\cite{9}}]{{\includegraphics[width = 1.32in]{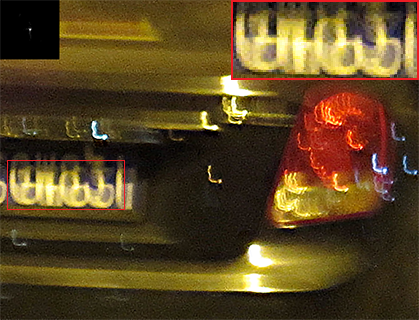}}}
\subfigure[Pan \textit{et al.} {\cite{13}}]{{\includegraphics[width = 1.32in]{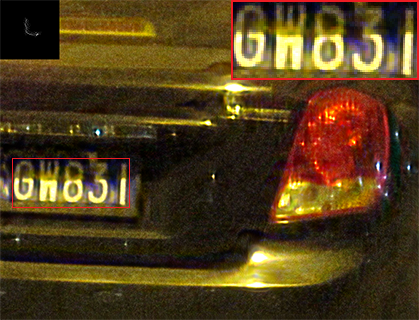}}}
\subfigure[Ours without PMP]{{\includegraphics[width = 1.32in]{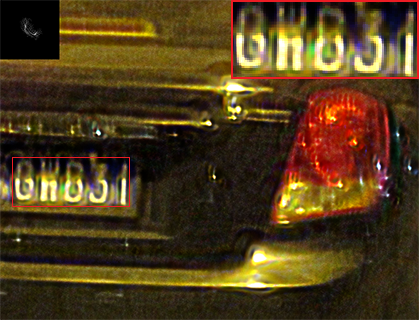}}}
\subfigure[Ours]{{\includegraphics[width = 1.32in]{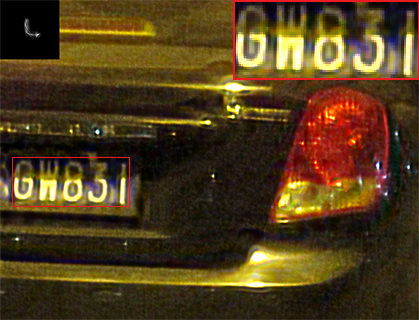}}}
\caption{Visual comparison on two text images deblurring.}
\label{figure14}
\end{figure*}

\textbf{Text image:}
The results of the compared algorithms on two text images are illustrated in Fig. \ref{figure14}.
Our algorithm performs comparably with the method \cite{13}.
When without using the PMP regularization,
our algorithm may fail to reconstruct the correct blur kernel
and yields a result with heavy ringing artifacts.

\begin{table}[!t]
\renewcommand\arraystretch{1.05}
\caption{Quantitative results on the dataset \cite{11}, including the average PSNR and average SSIM.}
\centering
\begin{tabular}{|l|c|c|}
\hline
 Method &  PSNR (dB) &  SSIM \\
\hline
Cho and Lee & 26.7548 & 0.8224\\
\hline
Cho \textit{et al.} & 17.9599 & 0.4922\\
\hline
Krishnan \textit{et al.} & 23.4366 & 0.7571\\
\hline
Levin \textit{et al.}  & 25.4989 & 0.8079\\
\hline
Xu and Jia & 29.1466 & 0.8553 \\
\hline
Pan \textit{et al.} & 30.2502 & 0.8587\\
\hline
Ours & \textbf{30.5660} & \textbf{0.8593}\\
\hline
\end{tabular}
\end{table}

\begin{figure*}[!t]
\centering
\subfigure[Blurred image]{{\includegraphics[width = 1.32in]{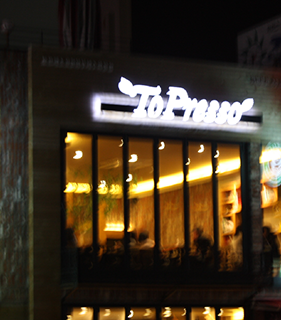}}}
\subfigure[Hu \textit{et al.} {\cite{15}}]{{\includegraphics[width = 1.32in]{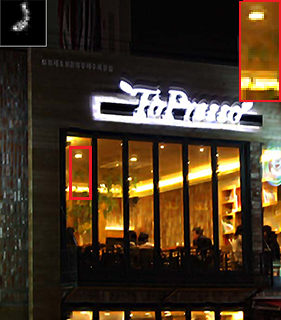}}}
\subfigure[Pan \textit{et al.} {\cite{13}}]{{\includegraphics[width = 1.32in]{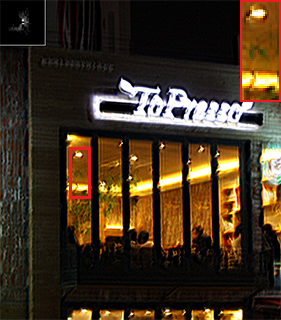}}}
\subfigure[Ours without PMP]{{\includegraphics[width = 1.32in]{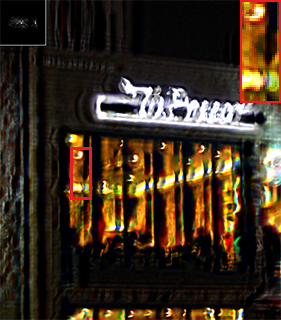}}}
\subfigure[Ours]{{\includegraphics[width = 1.32in]{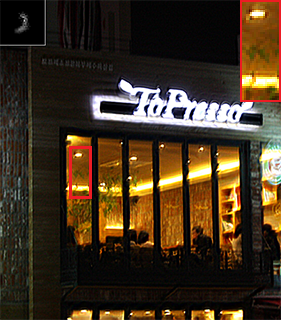}}}
\caption{Visual comparison on a low-light image.}
\label{figure15}
\end{figure*}

\textbf{Low-light image:}
Low-light images usually cannot be well handled by most deblurring methods.
A main reason is that low-light images often have saturated pixels
which interfere with the kernel estimation process \cite{15,16}.
Fig. \ref{figure15} presents the results on a low-light image.
The state-of-the-art low-light image deblurring method \cite{15} is used in the comparison.
Compared with the method \cite{15} specifically designed for low-light images,
our method gives a comparable result.

\subsection{Computational Complexity}

Finally, we compare the computational complexity of our algorithm ($r = 0.025 \cdot mean(m,n)$) with
those of the algorithms \cite{9, 13, 17}. Our algorithm without PMP and our algorithm with different patch size $r \in \{4,8,16\}$ are also compared.
As explained in Section IV, the proposed  Algorithm 1 would be more efficient than the traditional half quadratic splitting (HQS) algorithm, e.g. Algorithm 3.
To substantiate this, the runtime of the HQS algorithm using the PMP (Algorithm 3) is also compared (with $r\! = \!0.025\! \cdot\! mean(m,n)$).
The simulations are conducted under Windows 10
on a desktop PC with an Intel Core i7-4790 CPU at 3.6 GHz with 16 GB RAM.
For our method and the methods \cite{13,17},
the runtime of the non-blind deblurring step is included in the results.
Among these algorithms, one can observe from Table VI that the algorithm developed
by Xu \textit{et al.} \cite{9} with C++ implementation is the fastest.
However, in some cases, its restoration quality is inferior to
our algorithm as illustrated earlier in the above figures.
Our algorithm is much faster than the algorithms \cite{13,17}.
Compared with the algorithm of Pan \textit{et al.} \cite{13},
our algorithm is more than an order of magnitude faster.
Note that, the algorithm \cite{13}
can be accelerated in the dark-channel computation step as mentioned by the authors.
The result of this algorithm presented here is obtained
by the code provided by the authors at their website,
which is implemented without such acceleration.

\begin{table}[!t]
\renewcommand\arraystretch{1.05}
\caption{Runtime comparison in seconds (the kernel size is fixed at $51\times51$ for each algorithm).}
\centering
\begin{tabular}{|l|c|c|c|}
\hline
 Method    &  $256\!\times\!256$  &   $512\!\times\!512$ &  $800\!\times\!800$ \\
\hline
  Xu \textit{et al.} (C++)   & 1.05 & 2.43 & 5.35 \\
\hline
  Levin \textit{et al.} (Matlab)   & 155.9 & 657.8 & 1598.6 \\
\hline
  Pan \textit{et al.} (Matlab)   & 162.5 & 548.6 & 1261.3 \\
\hline
  HQS (Alg. 3) (Matlab)   & 107.3 & 401.2  &  938.8\\
\hline
  Ours w/o PMP (Matlab)   & 5.59 & 14.51  &  35.03\\
\hline
  Ours ($r=4$) (Matlab)   & 38.88 & 132.7  &  295.3\\
\hline
 Ours ($r=8$) (Matlab)   & 16.39 & 55.73  &  134.8\\
\hline
 Ours ($r=16$) (Matlab)   & 11.02  &  41.69  &  94.74 \\
\hline
 Ours ($r\! =\! 0.025 m$) (Matlab) & 18.61 & 44.36 & 95.55\\
\hline
\end{tabular}
\end{table}

\section{Conclusion}

This work proposed a local minimal intensity
based prior, namely PMP, and an improved algorithm for blind image deblurring.
The prior is simple yet effective in discriminating between clear and blurred images.
Rather than directly using the half quadratic splitting algorithm,
the new algorithm flexibly imposes sparsity inducing on the PMP
in the deblurring procedure under the MAP framework.
{Particularly, it avoids non-rigorous approximate solution in existing algorithms
in jointly handling multiple non-explicit priors,
while being much more efficient.}
Extensive experiments on both natural and specific images demonstrated that
it not only can achieve state-of-the-art deblurring quality,
but also can improve the practical stability and computational efficiency substantially.
In brief, in terms of both the practical robustness and computational efficiency,
the proposed algorithm is superior to the compared algorithms in this work.

\end{document}